\pdfoutput=1
\documentclass[
    affil-it, 
    auth-lg, 
    british, 
    contents, 
    custom-packages={}, 
    references={references}, 
]{lib/preprint}

\newcommand{\bmomega}{\bm\omega}

\begin{document}

    \date{20th July 2022}

    \email{d.bielli@surrey.ac.uk,silvia.penati@mib.infn.it,dmitri.sorokin@pd.infn. it,m.wolf@surrey.ac.uk}

    \preprint{DMUS--MP--21/16}

    \title{Super Non-Abelian T-Duality}

    \author[a,b]{Daniele~Bielli}
    \author[a]{Silvia~Penati}
    \author[c]{Dmitri~Sorokin}
    \author[b]{Martin~Wolf}

    \affil[a]{Dipartimento di Fisica, Universit{\`a} degli studi di Milano--Bicocca,\\ and INFN, Sezione di Milano--Bicocca, Piazza della Scienza 3, 20126 Milano, Italy}
    \affil[b]{Department of Mathematics, University of Surrey,\\ Guildford GU2 7XH, United Kingdom}
    \affil[c]{INFN, Sezione di Padova, and Dipartimento di Fisica e Astronomia ``Galileo Galilei'', Universit{\`a} degli Studi di Padova, Via F.~Marzolo 8, 35131 Padova, Italy}

    \abstract{We analyse super non-Abelian T-duality for principal chiral models, symmetric space sigma models, and semi-symmetric space sigma models for general Lie supergroups. This includes T-duality along both bosonic and fermionic directions. As an example, we perform the explicit dualisation of the $\sfOSp(1|2)$ principal chiral model, and, whilst the target superspace of this model is a three-dimensional supergravity background, we find that its super non-Abelian T-dual falls outside the class of such backgrounds.}

    \acknowledgements{We gratefully acknowledge stimulating conversations with Antonio Grassi, Jan Gutowski, Yolanda Lozano, Anayeli Ramirez, Christian Saemann, Alessandro Torrielli, and Linus Wulff. We also thank the referee for useful comments. DB and SP were partially supported by Universit{\`a} degli studi di Milano-Bicocca, by the Italian Ministero dell'Universit{\`a} e della Ricerca (MUR), and by the Istituto Nazionale di Fisica Nucleare (INFN) through the research project `Gauge theories, Strings, Supergravity' (GSS). DS was partially supported by the INFN research project `String Theory and Fundamental Interactions' (STEFI), by the Spanish MICINN/FEDER (ERDF EU) grant PGC2018-095205-B-I00 and by the Basque Government Grant IT-979-16.}

    \datalicencemanagement{No additional research data beyond the data presented and cited in this work are needed to validate the research findings in this work. For the purpose of open access, the authors have applied a Creative Commons Attribution (CC BY) licence to any Author Accepted Manuscript version arising.}

    \begin{body}

        \section{Introduction}

        The original `radius $\to$ 1/radius' duality of string theory discovered in~\cite{Sathiapalan:1986zb,Kikkawa:1984cp} has evolved into a more general concept of T-duality. This is based on a very simple mechanism: two classically duality-equivalent string sigma models exist whenever we can formulate a first-order Lagrangian which provides a set of equations of motion and Bianchi identities. Then, on the one hand, when imposing the Bianchi identities, one recovers the original model, while on the other hand, when imposing the equations of motion, one obtains the dual model.

        Abelian T-duality~\cite{Buscher:1987sk,Buscher:1987qj}, which is present when the string target space geometry has an Abelian isometry, is an exact symmetry of string theory that maps string backgrounds into dual ones corresponding to the same conformal field theory. In other words, it relates two world-sheet non-linear sigma models with equivalent physical properties, both at the classical level as well as the quantum level~\cite{Rocek:1991ps,Giveon:1991jj}.

        Instead, non-Abelian T-duality~\cite{delaOssa:1992vci}\footnote{See also~\cite{Nappi:1979ig,Fridling:1983ha,Fradkin:1984ai} for previous results.}, which generalises the approach of~\cite{Rocek:1991ps} to string backgrounds that exhibit a bosonic non-Abelian group of isometries, does not represent an exact string symmetry~\cite{Giveon:1993ai}. Indeed, the non-Abelian T-dual model possesses a different, in general smaller, set of symmetries (local as well as non-local ones). Notable examples can be found in~\cite{delaOssa:1992vci,Alvarez:1993qi,Sfetsos:1994vz,Lozano:2011kb}. If supersymmetry is part of the game, the T-dual model is, in general, less or non-supersymmetric, as can be inferred from studying the Killing spinor equations after dualisation~\cite{Sfetsos:2010uq}. Nevertheless, when starting with a string theory background with a semi-simple group of isometries, the T-dual model usually remains a string theory background essentially because non-Abelian T-duality preserves conformal invariance in this case.\footnote{In the case of non-semi-simple groups, this is no longer true~\cite{Gasperini:1993nz,Gasperini:1994du} due to the appearance of a mixed gauge-gravitational anomaly~\cite{Giveon:1993ai,Alvarez:1994np}. As shown in~\cite{Fernandez-Melgarejo:2017oyu,Hong:2018tlp}, such backgrounds are solutions to generalised supergravity field equations~\cite{Arutyunov:2015mqj,Wulff:2016tju}.} Therefore, non-Abelian T-duality can fruitfully be used as a string/supergravity solution generating technique~\cite{Sfetsos:2010uq,Lozano:2011kb}.

        Abelian fermionic T-duality has been formulated in~\cite{Berkovits:2008ic} as a natural generalisation of the Abelian bosonic T-duality to sigma models which possess a superisometry generated by an `Abelian' $Q$ supercharge ($Q^2=0$). The T-self-duality~\cite{Ricci:2007eq,Beisert:2008iq} of string theory on AdS$_5\times S^5$ under a combination of four bosonic and eight fermionic Abelian T-duality transformations provides a dual string explanation of the duality between maximally-helicity-violating amplitudes and light-like Wilson loops in $\caN=4$ supersymmetric Yang--Mills theory (see~\cite{Alday:2008yw} for a nice review). Since under T-duality ordinary superconformal invariance acting on Wilson loops is mapped to dual superconformal invariance of scattering amplitudes, which ultimately can be thought as being generated by some non-local current, the investigation of T-self-duality is intimately linked to integrability. For the AdS$_5\times S^5$ background this has been discussed in~\cite{Ricci:2007eq,Beisert:2008iq}.

        More generally, T-self-duality under combinations of bosonic and fermionic T-dualities has been proved for string sigma models on AdS$_d\times S^d \times M^{10-2d}$ in~\cite{Abbott:2015mla}. Furthermore, whilst T-self-duality for the AdS$_4\times\IC P^3$ string sigma model is also expected to hold, its proof remains an open problem. In particular, the existence of T-self-duality for this background is supported by the amplitudes/Wilson loops duality in three-dimensional Aharony--Bergman--Jafferis--Maldacena theory~\cite{Bianchi:2011rn,Bianchi:2011dg,Bianchi:2011fc,Basso:2018tif} as well as by the integrability emerging on the string theory sides~\cite{Arutyunov:2008if,Stefanski:2008ik,Gomis:2008jt,Sorokin:2010wn} and on the gauge theory side~\cite{Minahan:2008hf,Bak:2008cp}. However, the proof of T-self-duality in this case is hindered by the appearance of singularities in the T-duality transformations of the fields~\cite{Grassi:2009yj,Adam:2010hh,Bakhmatov:2010fp,Dekel:2011qw,Colgain:2016gdj}, and a stringy explanation of the amplitudes/Wilson loops duality in three dimensions remains open.

        Evidently, Abelian fermionic T-duality should be generalised to a non-Abelian setting when non-Abelian isometries form supergroups. This can be achieved by following a prescription similar to the one introduced in~\cite{delaOssa:1992vci} for purely bosonic non-Abelian isometries. However, the procedure of integrating out degrees of freedom to obtain the dual action appears more complicated because of the nilpotent nature of the fermionic coordinates involved in the dualisation~\cite{Grassi:2011zf}. This is one of the reasons why the studies in this direction have been rather limited so far. Dualisation of fermionic directions has been discussed somewhat implicitly in connection with $\lambda$-deformed sigma models~\cite{Borsato:2016zcf}, introduced in~\cite{Sfetsos:2013wia,Hollowood:2014rla,Hollowood:2014qma}, and with the construction of Yang--Baxter-deformed sigma models as non-Abelian T-duals of deformed coset superspaces~\cite{Borsato:2017qsx,Borsato:2018idb,vanTongeren:2019dlq}. In particular, it was proved in~\cite{Borsato:2018idb} that for a type II Green--Schwarz superstring, non-Abelian T-duality always produces a dual background which again satisfies the torsion constraints. This was shown under the assumption that one can pick the background $B_2$-field that is invariant under the dualised isometries. This is possible e.g.~for the Ramond--Ramond backgrounds described by semi-symmetric coset superspaces, but, in general, it is not the case for Neveu--Schwarz--Neveu--Schwarz backgrounds. In ten-dimensional supergravity, a non-Abelian version of fermionic T-duality has been formulated in~\cite{Astrakhantsev:2021rhj} but is limited, however, to fermionic isometries whose algebra closes on Abelian bosonic isometries. These lead to non-geometric solutions to double field theory subject to an extra constraint whose origin remains unclear. Generalisations of T-duality to supermanifolds and super Lie groups has also been explored in the context of Poisson--Lie T-duality~\cite{Eghbali:2009cp,Eghbali:2011su,Eghbali:2012sr,Eghbali:2013bda,Eghbali:2014coa}, originally introduced  as an extension of standard bosonic dualisation~\cite{Klimcik:1995ux,Klimcik:1995jn}.

        Despite these efforts, a clear picture of non-Abelian T-duality for supergroups is still missing. We shall refer to this as \emph{super non-Abelian T-duality} in the following. One of the main open problems is to understand better the interplay between bosonic and fermionic non-Abelian T-duality. In particular, it would be interesting to understand whether the combination of fermionic and bosonic non-Abelian T-dualities can provide a more extensive tool for generating new string/supergravity solutions.

        Inspired by these questions, in this paper we shall carry out a systematic study of super non-Abelian T-duality, when the dualisation is generically done along the directions corresponding to a non-Abelian subsupergroup of the supergroup of isometries. Using the current algebra approach, we perform in full generality the super non-Abelian T-duality of principal chiral models on supergroup manifolds, as well as of symmetric and semi-symmetric supercoset models. For all these models, we construct the T-dual action and identify the corresponding equations of motion.

        A well-known feature of bosonic T-duality is that integrability is preserved. Indeed, if the original model possesses a flat conserved current, thus allowing for the existence of a Lax connection, the conservation equation and the flatness condition get exchanged under dualisation, and (some of the) local charges of the original model are mapped to non-local charges of the T-dual model, and vice versa.\footnote{For an earlier discussion of dualities and integrability of two-dimensional sigma models, see e.g.~\cite{Pohlmeyer:1975nb,Luscher:1977rq,Eichenherr:1979ci,Eichenherr:1979hz,Eichenherr:1981sk,Ivanov:1987yv}.} This continues to be the case under super non-Abelian T-duality, and the T-dual Lax connection is formally of the same form as the original one.

        The three classes of models under investigation include physically relevant sigma models describing string/supergravity backgrounds, directly formulated in superspace. In order to address the question whether super non-Abelian T-duality always gives rise to new, non-equivalent supergravity solutions, it is important to investigate whether the T-dual geometry is compatible with superspace supergravity constraints.

        We investigate this issue in one particular example, the $\sfOSp(1|2)$ principal chiral model which describes a three-dimensional anti-de Sitter supergravity geometry. For ordinary three-dimensional anti-de Sitter geometry realised as an $\sfSL(2,\IR)$ principal chiral model, it has been shown~\cite{Alvarez:1993qi} that the T-dualisation of the whole $\sfSL(2,\IR)$ isometry group leads to a dual geometry that describes a black hole that is asymptotically anti-de Sitter, though the three-dimensional matter-coupled gravity in which this black hole may arise as a solution is unknown yet. Aimed at investigating whether this black hole has a conventional supersymmetric generalisation, we study super non-Abelian T-duality of the $\sfOSp(1|2)$ model whose geometry is that of the $\caN=1$ AdS$_3$ superspace. Dualising either the whole $\sfOSp(1|2)$ supergroup or its maximal bosonic $\sfSL(2,\IR)$ subgroup, we argue that the T-dual model does not satisfy the three-dimensional supergravity constraints~\cite{Kuzenko:2011xg,Kuzenko:2011rd,Kuzenko:2012bc,Kuzenko:2013uya}, thus falling outside the class of supergravity backgrounds. The reason for this is the fact that the super non-Abelian-T-duality procedure gives rise to dual supervielbeins and a $B_2$-field with a complicated space-time dependence that prevents them to satisfy the three-dimensional supergravity constraints (even after any possible redefinition of supervielbeins and connection). This seems to be a first example in which super non-Abelian T-duality does not generate supergravity solutions. Whether this failure is a peculiarity of this model or arises more generally remains an open question that we plan to address in a near future. We should note that this result does not contradict the result of~\cite{Borsato:2018idb} since, as will be explained in detail below, the $\sfOSp(1|2)$ sigma model under consideration does not describe a (three-dimensional) Green--Schwarz superstring.

        This paper is organised as follows. In \cref{sec:preliminaries}, after a brief review of principal chiral models and supercosets, we summarise the general procedure for dualising (part of) their superisometries that we are going to apply in the rest of the paper. This is the direct generalisation of the procedure for purely bosonic non-Abelian T-duality~\cite{Rocek:1991ps,delaOssa:1992vci} based on gauging the subsupergroup involved in the dualisation. In \cref{sec:TDualityPCM}, we focus on the super non-Abelian T-duality of principal chiral models. After deriving the T-dual action in general and studying its symmetries, we investigate in detail the $\sfOSp(1|2)$ principal chiral model describing the super AdS$_3$ geometry. In particular, we argue that the target superspace of the T-dual model obtained by dualising either the whole supergroup or its maximal bosonic subgroup does not satisfy the three-dimensional supergravity constraints. In \cref{sec:cosets}, we give preliminary results regarding the application of super non-Abelian T-duality to supercosets. Since for these models the dualisation procedure can be affected by the non-invertibility of the gauge equations of motion required to remove the original coordinates in favor of a consistent set of dual ones, we cannot determine the T-dual action in general. At this point, a case-by-case study would be necessary which is, however, beyond the scope of the present paper. Finally, in \cref{sec:conclusions}, we close by collecting some comments and a list of possible future directions. Several appendices follow, which contain some technical details and a brief summary of three-dimensional supergravity (see \cref{app:conformalSUGRA}).

        \section{Preliminaries}\label{sec:preliminaries}

        We begin with a short review of the classes of two-dimensional sigma models which will be considered in the present paper. This primarily helps us fixing notation and conventions.

        We consider a generic Lie supergroup $\sfG$ with Lie superalgebra $\frg$. We denote the number of bosonic and fermionic generators of $\frg$ by $n_\ttb$ and $n_\ttf$, respectively. In addition, we assume that we are given an inner product $\inner{-}{-}$ on $\frg$, that is, a non-degenerate Ad-invariant graded-symmetric bilinear form.

        Generally, to define two-dimensional sigma models involving $\sfG$, we consider elements $g\in\scC^\infty(\Sigma,\sfG)$ where $\scC^\infty(\Sigma,\sfG)$ is the set of smooth maps from a two-dimensional Lorentzian manifold $\Sigma$, the world-sheet, to $\sfG$. We then introduce the (pull-back to $\Sigma$ via $g$ of the) Maurer--Cartan form on $\sfG$ by
        \begin{equation}\label{eq:flatCurrent}
            j\ \coloneqq\ g^{-1}\rmd g\ \in\ \Omega^1(\Sigma,\frg)
        \end{equation}
        and which satisfies the Maurer--Cartan equation
        \begin{equation}\label{eq:MCEquation}
            \rmd j+\tfrac12[j,j]\ =\ 0~.
        \end{equation}
        Here, `$\rmd$' is the exterior derivative on $\Sigma$ and $\Omega^p(\Sigma,\frg)$ are the $\frg$-valued differential $p$-forms on $\Sigma$. In addition, the Lie bracket $[-,-]$ of $\frg$ is extended to $\Omega^p(\Sigma,\frg)$ in the usual way.

        The current in~\eqref{eq:flatCurrent} is invariant under the global (left) $\sfG$-action
        \begin{subequations}\label{eq:GAction}
            \begin{equation}\label{eq:leftGAction}
                g\ \mapsto\ g^{-1}_0g
                \eforall
                g_0\ \in\ \sfG~,
            \end{equation}
            whereas under the global (right) $\sfG$-action
            \begin{equation}\label{eq:rightGAction}
                g\ \mapsto\ g g_0
                \eforall
                g_0\ \in\ \sfG~,
            \end{equation}
        \end{subequations}
        it transforms adjointly $j\mapsto g_0^{-1}jg_0$.

        In the following, we shall often use the notation
        \begin{equation}\label{eq:covariantDerivative}
            \nabla_\omega\ \coloneqq\ \rmd+[\omega,-]
            \eand
            F_\omega\ \coloneqq\ \rmd\omega+\tfrac12[\omega,\omega]
            \efor
            \omega\ \in\ \Omega^1(\Sigma,\frg)~.
        \end{equation}
        Furthermore, if `$\star$' denotes the Hodge star operator on $\Sigma$ with respect to the world-sheet metric, we have $\star^2=\id$ on $\Omega^1(\Sigma,\frg)$. Therefore, we have the decomposition $\Omega^1(\Sigma,\frg)\cong\Omega^1_+(\Sigma,\frg)\oplus\Omega^1_-(\Sigma,\frg)$ into self-dual and anti-self-dual $\frg$-valued one-forms on $\Sigma$. We shall make use of the projectors
        \begin{equation}\label{eq:ASDProjectors}
            P^\pm\,:\,\Omega^1(\Sigma,\frg)\ \rightarrow\ \Omega^1_\pm(\Sigma,\frg)
            \ewith
            P^\pm\ \coloneqq\ \tfrac12(\id\,\pm\,\star)~.
        \end{equation}

        \subsection{Principal chiral model}\label{sec:PCM}

        \paragraph{Sigma model action.}
        The principal chiral model on $\sfG$ is defined by the action
        \begin{equation}\label{eq:PCMAction}
            S\ \coloneqq\ \tfrac12\int_\Sigma\inner{j}{\star j}~,
        \end{equation}
        where $j$ is the current in~\eqref{eq:flatCurrent}, the wedge product is understood and `$\star$' is the Hodge star operator on $\Sigma$ with respect to the world-sheet metric. The model describes the dynamics of $n_\ttb$ bosonic and $n_\ttf$ fermionic degrees of freedom. The equation of motion following from~\eqref{eq:PCMAction} is
        \begin{equation}\label{eq:EOMPCM}
            \rmd{\star j}\ =\ 0~,
        \end{equation}
        and together with the Maurer--Cartan equation~\eqref{eq:MCEquation}, this forms what is known as the first-order system of the model.

        \paragraph{Noether symmetries.}
        The action~\eqref{eq:PCMAction} possesses a global $(\sfG\times\sfG)$-symmetry, where, in our conventions, the first factor corresponds to the global left $\sfG$-action~\eqref{eq:leftGAction} and the second factor to the global right $\sfG$-action~\eqref{eq:rightGAction}. It is not too difficult to see that the corresponding Noether currents are
        \begin{equation}\label{eq:NoetherPCM}
            L_\rmN\ \coloneqq\ gjg^{-1}
            \eand
            R_\rmN\ \coloneqq\ j~.
        \end{equation}

        \paragraph{Lax connection.}
        Importantly, the first-order system, consisting of~\eqref{eq:EOMPCM} and~\eqref{eq:MCEquation}, is equivalent to
        \begin{subequations}\label{eq:PCMFlatnessCondition}
            \begin{equation}\label{eq:PCMFlatness}
                \rmd J(z)+\tfrac12[J(z),J(z)]\ =\ 0~,
            \end{equation}
            where
            \begin{equation}\label{eq:PCMLax}
                J(z)\ \coloneqq\ -\tfrac14(z- z^{-1})^2\,j +\tfrac14(z^2-z^{-2})\,{\star j}
            \end{equation}
        \end{subequations}
        is called the Lax connection with $z$ a complex spectral parameter~\cite{Zakharov:1973pp}. To derive the first-order system from~\eqref{eq:PCMFlatnessCondition}, we note that $\star^2=\id$ and $\rho\wedge{\star\sigma}=-{\star\rho}\wedge\sigma$ for $\rho,\sigma\in\Omega^1(\Sigma,\frg)$. As is well-known, a flat Lax connection always yields an infinite number of conservation laws~\cite{Luscher:1977rq}, thus ensuring the integrability of the model (for a nice recent review, see~\cite{Torrielli:2016ufi}).

        \subsection{Symmetric space sigma model}\label{sec:Symmetric}

        \paragraph{Lie algebra decomposition.}
        We now consider sigma models defined on coset superspaces $\sfG/\sfH$ where $\sfH$ is a Lie subsupergroup of $\sfG$ which arises as the fixed point set of an automorphism of $\sfG$ of order 2. Such coset spaces are known as symmetric spaces. At the level of the Lie superalgebras, this means that the Lie superalgebra $\frg$ of $\sfG$ decomposes as
        \begin{subequations}\label{eq:symmetricSpaceDecomposition}
            \begin{equation}
                \frg\ \cong\ \frh\oplus\frm
                \ewith
                \frh\ \coloneqq\ \underbrace{\tfrac12(1+\sigma)}_{\eqqcolon\,P_\frh}(\frg)
                \eand
                \frm\ \coloneqq\ \underbrace{\tfrac12(1-\sigma)}_{\eqqcolon\,P_\frm}(\frg)~,
            \end{equation}
            where $\frh$ is the Lie superalgebra of $\sfH$ and $\sigma:\frg\rightarrow\frg$ is an automorphism of $\frg$ with $\sigma^2=1$. Then,
            \begin{equation}\label{eq:symmetricSpaceConditions}
                [\frh,\frh]\ \subseteq\ \frh~,
                \quad
                [\frm,\frh]\ \subseteq\ \frm~,
                \eand
                [\frm,\frm]\ \subseteq\ \frh~.
            \end{equation}
        \end{subequations}
        Elements of $\frh$ are said to be of homogeneity $0$ and elements of $\frm$ of homogeneity $2$, respectively and we denote the homogeneity of a homogeneous element by $|-|$. Furthermore, we assume that the inner product $\inner{-}{-}$ is compatible with the decomposition~\eqref{eq:symmetricSpaceDecomposition}, that is, we take $\inner{U}{V}=0$ for $|U|+|V|\neq 0$ mod $4$ for any two homogeneous elements $U,V\in\frg$.

        \paragraph{Sigma model action.}
        In order to formulate the sigma model action, we first observe that under the decomposition~\eqref{eq:symmetricSpaceDecomposition} the current $j=g^{-1}\rmd g$ decomposes as
        \begin{equation}\label{eq:symmetricSpaceFlatCurrent}
            \begin{gathered}
                j\ =\ A+m~,
                \\
                A\ \coloneqq\ P_\frh(j)\ \in\ \Omega^1(\Sigma,\frh)
                \eand
                m\ \coloneqq\ P_\frm(j)\ \in\ \Omega^1(\Sigma,\frm)~.
            \end{gathered}
        \end{equation}
        Under the local (right) $\sfH$-action
        \begin{equation}\label{eq:HAction}
            g\ \mapsto\ gh
            \eforall
            h\ \in\ \scC^\infty(\Sigma,\sfH)
        \end{equation}
        the $A$ component in~\eqref{eq:symmetricSpaceFlatCurrent} transforms as a connection one-form, $A\mapsto h^{-1}A h+h^{-1}\rmd h$, whereas $m$ as an endomorphism one-form, $m\mapsto h^{-1}m h$. Therefore, the sigma model action
        \begin{equation}\label{eq:symmetricSpaceAction}
            S\ \coloneqq\ \tfrac12\int_\Sigma\inner{m}{\star m}
        \end{equation}
        is invariant under local transformations~\eqref{eq:HAction}. The associated equation of motion reads
        \begin{equation}\label{eq:EOMSymmetricSpace}
            \nabla_A{\star m}\ =\ 0~,
        \end{equation}
        where $\nabla_A$ has been defined in~\eqref{eq:covariantDerivative}. This model describes the dynamics of $n_\ttb=n_\ttb^\sfG-n_\ttb^\sfH$ bosonic and $n_\ttf=n_\ttf^\sfG-n_\ttf^\sfH$ fermionic degrees of freedom, with $n_\ttb^\sfG,n_\ttb^\sfH$ and $n_\ttf^\sfG,n_\ttf^\sfH$ being the number of bosonic and fermionic generators of $\sfG$ and $\sfH$, respectively.

        \paragraph{Noether symmetries.}
        The action~\eqref{eq:symmetricSpaceAction} is invariant under the global (left) $\sfG$-action~\eqref{eq:leftGAction}. It is easy to check that the associated Noether current is given by
        \begin{equation}\label{eq:NoetherSymmetricSpace}
            J_\rmN\ \coloneqq\ gmg^{-1}~.
        \end{equation}

        \paragraph{Lax connection.}
        The Maurer--Cartan equation~\eqref{eq:MCEquation} decomposes under~\eqref{eq:symmetricSpaceDecomposition} as
        \begin{equation}\label{eq:MCSymmetricSpace}
            F_A+\tfrac12[m,m]\ =\ 0
            \eand
            \nabla_A m\ =\ 0
        \end{equation}
        and together with the equation of motion~\eqref{eq:EOMSymmetricSpace} they constitute the first-order system of a symmetric space sigma model. Importantly, this first-order system is equivalent to
        \begin{subequations}
            \begin{equation}
                \rmd J(z)+\tfrac12[J(z),J(z)]\ =\ 0~,
            \end{equation}
            where
            \begin{equation}\label{eq:symmetricSpaceLax}
                J(z)\ \coloneqq\ A +\tfrac12(z^2+z^{-2})\,m-\tfrac12(z^2-z^{-2})\,{\star m}
            \end{equation}
        \end{subequations}
        is the Lax connection with $z$ a complex spectral parameter.

        Evidently, this Lax connection depends on the choice of the coset representative. However, in order to generate an infinite tower of conserved charges which are manifestly gauge invariant, it is preferable to consider a Lax connection that is invariant under $\sfH$-gauge transformations. This requirement is satisfied by
        \begin{equation}
            J'(z)\ \coloneqq\ g\big(J(z)-J(1)\big)g^{-1}\ =\ g\big(J(z)-j\big)g^{-1}\ =\ gJ(z)g^{-1}+g\rmd g^{-1}~.
        \end{equation}
        Furthermore, upon defining $z'\coloneqq-\log(z)$ and expanding this expression around $z'=0$, we obtain
        \begin{equation}
            J'(z(z'))\ =\ {\star L_\rmN}\,z'+\caO(z'^2)~,
        \end{equation}
        where the coefficent $L_\rmN$ is the Noether current~\eqref{eq:NoetherSymmetricSpace}. Thus, at order $z'$ the flatness of $J'$ implies the conservation of the Noether current.

        \subsection{Semi-symmetric space sigma model}\label{sec:Semisymmetric}

        \paragraph{Lie algebra decomposition.}
        Let us now come to sigma models on semi-symmetric spaces, that is coset superspaces $\sfG/\sfH$ where $\sfH$ is a Lie subsupergroup of $\sfG$ which arises as the fixed point set of an automorphism of $\sfG$ of order 4. This means that the Lie superalgebra $\frg$ of $\sfG$ decomposes as
        \begin{subequations}\label{eq:semiSymmetricSpaceDecomposition}
            \begin{equation}
                \frg\ \cong\ \frh\oplus\frp\oplus\frm\oplus\frq
            \end{equation}
            with
            \begin{equation}
                \begin{gathered}
                    \frh\ \coloneqq\ \underbrace{\tfrac12(1+\sigma+\sigma^2+\sigma^4)}_{\eqqcolon\,P_\frh}(\frg)~,
                    \quad
                    \frm\ \coloneqq\ \underbrace{\tfrac12(1-\sigma+\sigma^2-\sigma^3)}_{\eqqcolon\,P_\frm}(\frg)~,
                    \\
                    \frp\ \coloneqq\ \underbrace{\tfrac12(1-\rmi\sigma-\sigma^2+\rmi\sigma^4)}_{\eqqcolon\,P_\frp}(\frg)~,
                    \quad
                    \frq\ \coloneqq\ \underbrace{\tfrac12(1+\rmi\sigma-\sigma^2-\rmi\sigma^3)}_{\eqqcolon\,P_\frq}(\frg)~,
                \end{gathered}
            \end{equation}
            where $\frh$ is the Lie superalgebra of $\sfH$ and $\sigma:\frg\rightarrow\frg$ is an automorphism of $\frg$ with $\sigma^4=1$ and
            \begin{equation}\label{eq:semiSymmetricSpaceConditions}
                \begin{gathered}
                    [\frh,\frh]\ \subseteq\ \frh~,
                    \quad
                    [\frp,\frh]\ \subseteq\ \frp~,
                    \quad
                    [\frm,\frh]\ \subseteq\ \frm~,
                    \quad
                    [\frq,\frh]\ \subseteq\ \frq~,
                    \quad
                    \\
                    [\frp,\frp]\ \subseteq\ \frm~,
                    \quad
                    [\frm,\frp]\ \subseteq\ \frq~,
                    \quad
                    [\frq,\frp]\ \subseteq\ \frh~,
                    \\
                    [\frm,\frm]\ \subseteq\ \frh~,
                    \quad
                    [\frq,\frm]\ \subseteq\ \frp~,
                    \\
                    [\frq,\frq]\ \subseteq\ \frm~.
                \end{gathered}
            \end{equation}
        \end{subequations}
        Elements of $\frh$ are said to be of homogeneity $0$, elements of $\frp$ of homogeneity $1$, elements of $\frm$ of homogeneity $2$, and elements of $\frq$ of homogeneity $3$, respectively. Furthermore, we assume that the inner product $\inner{-}{-}$ is compatible with the decomposition~\eqref{eq:semiSymmetricSpaceDecomposition}, that is, we take $\inner{U}{V}=0$ for $|U|+|V|\neq 0$ mod $4$ for any two homogeneous elements $U,V\in\frg$.

        \paragraph{Sigma model action.}
        Under the decomposition~\eqref{eq:semiSymmetricSpaceDecomposition} the current~\eqref{eq:flatCurrent} decomposes as
        \begin{equation}\label{eq:semiSymmetricSpaceFlatCurrent}
            \begin{gathered}
                j\ =\ A+p+m+q~,
                \\
                A\ \coloneqq\ P_\frh(j)\ \in\ \Omega^1(\Sigma,\frh)~,
                \quad
                m\ \coloneqq\ P_\frm(j)\ \in\ \Omega^1(\Sigma,\frm)~,
                \\
                p\ \coloneqq\ P_\frp(j)\ \in\ \Omega^1(\Sigma,\frp)~,
                \quad
                q\ \coloneqq\ P_\frq(j)\ \in\ \Omega^1(\Sigma,\frq)~.
            \end{gathered}
        \end{equation}
        To formulate the sigma model action, we consider that under the local (right) $\sfH$-action~\eqref{eq:HAction}  the component $A$ of $j$ transforms as a connection one-form, $h\mapsto h^{-1}Ah+h^{-1}\rmd h$, whereas $\{p,m,q\}$ as endomorphism one-forms, $\{p,m,q\}\mapsto h^{-1}\{p,m,q\}h$ for all $h\in\scC^\infty(\Sigma,\sfH)$. The associated sigma model action is then the Green--Schwarz-like action~\cite{Metsaev:1998it,Berkovits:1999zq,Roiban:2000yy}
        \begin{equation}\label{eq:semiSymmetricSpaceAction}
            S\ \coloneqq\ \tfrac12\int_\Sigma\inner{m}{\star m}+\tfrac12\int_\Sigma\inner{p}{q}~.
        \end{equation}

        The equations of motion following from~\eqref{eq:semiSymmetricSpaceAction} are
        \begin{equation}\label{eq:EOMSemiSymmetricSpace}
            \begin{aligned}
                \nabla_A {\star m}-\tfrac12[p,p]+\tfrac12[q,q]\ &=\ 0~,
                \\
                [m,p+{\star p}]\ &=\ 0~,
                \\
                [m,q-{\star q}]\ &=\ 0~.
            \end{aligned}
        \end{equation}
        As in the symmetric space case, the model describes the dynamics of $n_\ttb$ bosonic and $n_\ttf$ fermionic degrees of freedom with $n_\ttb=n_\ttb^\sfG-n_\ttb^\sfH$ and $n_\ttf=n_\ttf^\sfG-n_\ttf^\sfH$.

        \paragraph{Noether symmetries.}
        One can check that
        \begin{equation}\label{eq:NoetherSemiSymmetricSpace}
            J_\rmN\ \coloneqq\ g\big(m-\tfrac12{\star(p-q)}\big)g^{-1}
        \end{equation}
        is the Noether current associated with the global (left) $\sfG$-action~\eqref{eq:leftGAction}.

        \paragraph{Lax connection.}
        The Maurer--Cartan equation~\eqref{eq:MCEquation} decomposes under~\eqref{eq:semiSymmetricSpaceDecomposition} as
        \begin{equation}
            \begin{aligned}
                F_A+\tfrac12[m,m]+[p,q]\ &=\ 0~,
                \\
                \nabla_A p+[m,q]\ &=\ 0~,
                \\
                \nabla_A m+\tfrac12[p,p]+\tfrac12[q,q]\ &=\ 0~,
                \\
                \nabla_A q+[m,p]\ &=\ 0~,
            \end{aligned}
        \end{equation}
        and together with the equations of motion~\eqref{eq:EOMSemiSymmetricSpace} they constitute the first-order system of a semi-symmetric space sigma model. Importantly, this first-order system is equivalent to
        \begin{subequations}
            \begin{equation}
                \rmd J(z)+\tfrac12[J(z),J(z)]\ =\ 0~,
            \end{equation}
            where
            \begin{equation}\label{eq:semiSymmetricSpaceLax}
                J(z)\ \coloneqq\ A+zp+\tfrac12(z^2+z^{-2})\,m+z^{-1}q-\tfrac12(z^2-z^{-2})\,{\star m}
            \end{equation}
        \end{subequations}
        is the Lax connection with $z$ being a complex spectral parameter~\cite{Bena:2003wd}.\footnote{Generally, such Lax connections and flatness conditions arise for coset spaces that admit a $\IZ_m$-grading~\cite{Young:2005jv}.}

        As before, this Lax connection depends on the choice of coset representative but again, there is a Lax connection that is invariant under $\sfH$-gauge transformations,
        \begin{equation}
            J'(z)\ \coloneqq\ g\big(J(z)-J(1)\big)g^{-1}\ =\ g\big(J(z)-j\big)g^{-1}\ =\ gJ(z)g^{-1}+g\rmd g^{-1}~.
        \end{equation}
        Upon setting $z'\coloneqq-\log(z)$ and expanding this expression around $z'=0$, we obtain
        \begin{equation}
            J'(z(z'))\ =\ {\star L_\rmN}\,z'+\caO(z'^2)~,
        \end{equation}
        where now $L_\rmN$ is the Noether current~\eqref{eq:NoetherSemiSymmetricSpace}.

        \subsection{Topological deformations}\label{sec:topologicalDeformations}

        We may also consider deformations of the models~\eqref{eq:PCMAction},~\eqref{eq:symmetricSpaceAction}, and~\eqref{eq:semiSymmetricSpaceAction} that do neither alter the equations of motion nor the Noether symmetries, thus preserving integrability. Due to these properties, such type of deformations are referred to as \emph{topological}, and they have first been studied in the literature~\cite{Hoare:2016wsk,Borsato:2016pas,Borsato:2017qsx,Borsato:2018idb} in connection with Yang--Baxter deformations of two-dimensional sigma models~\cite{Klimcik:2002zj,Klimcik:2008eq,Delduc:2013qra,Kawaguchi:2014qwa}.

        \paragraph{Two-cocycles and derivations.}
        Let $\Omega\in H^2(\frg)$ be a Lie superalgebra two-cocycle (we may also restrict to a Lie subsuperalgebra of $\frg$)\footnote{The Whitehead lemma says that if $\frg$ is a finite-dimensional semi-simple Lie algebra, then $H^2(\frg)=0$. However, for Lie superalgebras, the Whitehead lemma does no longer hold in general. See~\cite{Scheunert:1997ks} and references therein on the cohomology for Lie superalgebras. See also~\cite{Catenacci:2020ybi,Cremonini:2021zwd} for some recent developments in cohomology theory for Lie superalgebras relevant to string theory/supergravity.}, that is, $\Omega\in\bigwedge^2\frg^*$ and it is subject to the cocycle condition
        \begin{equation}\label{eq:2-cocycle}
            \Omega(U,[V,W])+(-1)^{|U|(|V|+|W|)}\Omega(V,[W,U])+(-1)^{|W|(|U|+|V|)}\Omega(W,[U,V])\ =\ 0
        \end{equation}
        for all $U,V,W\in\frg$ where $|-|$ denotes the Gra{\ss}mann degree. By the Riesz representation theorem, there is a unique endomorphism $D:\frg\rightarrow\frg$ such that
        \begin{equation}\label{eq:definitionD}
           \Omega(U,V)\ =\ \inner{D(U)}{V}
           \eforall
           U,V\ \in\ \frg~.
        \end{equation}
        Whilst the graded anti-symmetry of $\Omega$ is equivalent to
        \begin{equation}\label{eq:DInnerProductAction}
           \inner{D(U)}{V}\ =\ -\inner{U}{D(V)}
           \eforall
           U,V\ \in\ \frg~,
        \end{equation}
        the cocycle condition~\eqref{eq:2-cocycle} is equivalent to
        \begin{equation}\label{eq:derivationD}
           D([U,V])\ =\ [D(U),V]+[U,D(V)]
           \eforall
           U,V\ \in\ \frg~,
        \end{equation}
        that is, $D$ is a derivation for the Lie bracket $[-,-]$. Furthermore, upon writing $g=\rme^V$ for $g\in\sfG$ and $V\in\frg$, we can extend $D$ to a left-invariant vector field on $\sfG$ by
        \begin{equation}\label{eq:extensionD}
            D(\rme^V)\ \coloneqq\ \rme^V\sum_{k=0}^\infty\frac{(-1)^k}{(k+1)!}\ad_V^k(D(V))
            \ewith
            \ad_V\ \coloneqq\ [V,-]~.
        \end{equation}
        It then follows that
        \begin{equation}\label{eq:AdActionD}
            D(gUg^{-1})\ =\ g\big(D(U)+\big[g^{-1}D(g),U\big]\big)g^{-1}
        \end{equation}
        for all $g\in\sfG$ and for all $U\in\frg$. See \cref{app:cocycles} for details. The Ad-invariance of $\inner{-}{-}$ then immediately yields
        \begin{equation}\label{eq:AdActionOmega}
           \Omega(gUg^{-1},gVg^{-1})\ =\ \Omega(U,V)+\inner{g^{-1}D(g)}{[U,V]}
        \end{equation}
        for all $g\in\sfG$ and for all $U,V\in\frg$. Moreover, we shall also make use of the notation
        \begin{equation}\label{eq:definitionCovariantD}
            D_A\ \coloneqq\ D+\ad_A
            \eforall
            A\ \in\ \frg~.
        \end{equation}
        Then,~\eqref{eq:AdActionD} implies that
        \begin{equation}\label{eq:gaugeTransformationCovariantD}
            D_{g^{-1}Ag+g^{-1}D(g)}(g^{-1}Ug)\ =\ g^{-1}D_A(U)g
        \end{equation}
        for all $g\in\sfG$ and for all $A,U\in\frg$. Note that to show this one uses that $g^{-1}D(g)=-D(g^{-1})g$ for all $g\in\sfG$ as directly follows from~\eqref{eq:extensionD}.

        \paragraph{Topological deformations.}
        Next, if we extend $\Omega$ and $D$ in the usual fashion to $\Omega^p(\Sigma,\frg)$, we can define a deformed model by
        \begin{equation}\label{eq:omegadef}
            S_\Omega\ \coloneqq\ S+\tfrac\zeta2\int_\Sigma\Omega(j,j)\ =\ S+\tfrac{\zeta}{2}\int_\Sigma\inner{D(j)}{j}~,
        \end{equation}
        where $S$ is any of the actions~\eqref{eq:PCMAction},~\eqref{eq:symmetricSpaceAction}, or~\eqref{eq:semiSymmetricSpaceAction} and $\zeta\in\IR$ is an arbitrary deformation parameter. In the case when considering coset superspaces $\sfG/\sfH$, that is, either~\eqref{eq:symmetricSpaceAction} or~\eqref{eq:semiSymmetricSpaceAction}, we need to require that the restriction $\Omega|_\frh$ of the two-cocyle $\Omega$ to the Lie superalgebra $\frh$ of $\sfH$ vanishes. This, in turn, ensures that the action~\eqref{eq:omegadef} is invariant under the $\sfH$-gauge transformations~\eqref{eq:HAction}. Note that the condition $\Omega|_\frh=0$ is equivalent to $D$ being of degree two, that is, $D|_\frh:\frh\rightarrow\frm$ and $D|_\frm:\frm\rightarrow\frh$ in the case of the symmetric space sigma models, and in the case of semi-symmetric space sigma models this gets augmented by $D|_\frp:\frp\rightarrow\frq$ and $D|_\frq:\frq\rightarrow\frp$.\footnote{Indeed, $\Omega|_\frh=0$ is equivalent to $D|_\frh:\frh\rightarrow\frm$, and the other conditions then follow from the derivation property of $D$ and the requirement of preserving the commutations relation~\eqref{eq:symmetricSpaceConditions} and~\eqref{eq:semiSymmetricSpaceConditions}.}

        Furthermore, it is not too difficult to check that the deformation is topological in the sense that the equations of motion for~\eqref{eq:omegadef} are the same as those of the undeformed model given by $S$. To show this, one makes use of the derivation property~\eqref{eq:derivationD} of $D$ and the Maurer--Cartan equation~\eqref{eq:MCEquation}. Likewise, it also follows that~\eqref{eq:omegadef} is invariant under the same symmetry transformation as the undeformed model with the same Noether currents. All this thus ensures that the deformed model has the same Lax connection as the undeformed model and hence, the deformation does not spoil integrability.\footnote{Another deformation one may consider is by using the Kac--Moody two-cocycle on the based loop algebra $L_0\frg$ given by $\int_\Sigma\int_0^1\rmd t\,\inner{\dder[J]{t}}{J}$ with $J:\Sigma\times S^1\rightarrow\frg$ with $\rmd J+\tfrac12[J,J]=0$ and $J(t=0)=J(t=1)=j$. Such a deformation also renders the theory integrable.}

        As mentioned before, we may consider the restriction $\Omega|_\frk$ of $\Omega$ to some Lie subsuperalgebra $\frk$ of $\frg$. If $\sfK$ denotes the Lie supergroup associated with $\frk$, we may consider the change $g\mapsto k^{-1}g$ for $k\in\scC^\infty(\Sigma,\sfK)$ and then deform any of the sigma model actions~\eqref{eq:PCMAction},~\eqref{eq:symmetricSpaceAction}, or~\eqref{eq:semiSymmetricSpaceAction} by adding $\frac\zeta2\int_\Sigma\Omega|_\frk(k^{-1}\rmd k,k^{-1}\rmd k)$ with $\Omega|_\frk\in H^2(\frk)$. In fact, under the assumption that $\Omega|_\frk$ is non-degenerate, homogeneous Yang--Baxter deformations have been shown in~\cite{Borsato:2016pas} to be equivalent to the non-Abelian T-dual of the deformed model when T-duality is performed, with $\Omega|_\frk^{-1}$ solving the (classical) Yang--Baxter equation and the deformation parameter given by $\zeta^{-1}$.

        \subsection{Super non-Abelian T-duality}

        We are now interested in studying non-Abelian T-duality of the sigma models from the preceeding sections. The general procedure that we shall apply is a straightforward generalisation of the well-known prescription of purely bosonic non-Abelian T-duality~\cite{Rocek:1991ps,delaOssa:1992vci} to supergroups, and we shall refer to it as \emph{super non-Abelian T-duality} in the following.

        \paragraph{Gauging.}
        Following~\cite{Rocek:1991ps,delaOssa:1992vci}, the first step to dualise one of the previously discussed sigma models is to gauge a Lie subsupergroup $\sfK$ of $\sfG$, corresponding to the directions that we wish to dualise. If $\frk$ denotes the Lie superalgebra of $\sfK$, the gauging is obtained by introducing a $\frk$-valued connection one-form $\omega\in\Omega^1(\Sigma,\frk)$. Consequently, the current~\eqref{eq:flatCurrent} generalises to
        \begin{equation}\label{eq:gaugedFlatCurrent}
            j_\omega\ \coloneqq\ g^{-1}\omega g+g^{-1}\rmd g
            \efor
            g\ \in\ \scC^\infty(\Sigma,\sfG)~.
        \end{equation}
        Evidently, $F_{j_\omega}=g^{-1}F_\omega g$ and so, the flatness of $j_\omega$ is equivalent to the flatness of $\omega$. Furthermore, $j_\omega$ is invariant under the local (left) $\sfK$-action
        \begin{equation}\label{eq:BGaugeTransformations}
            g\ \mapsto\ k^{-1}g
            \eand
            \omega\ \mapsto\ k^{-1}\omega k+k^{-1}\rmd k
            \eforall
            k\ \in\ \scC^\infty(\Sigma,\sfK)~.
        \end{equation}

        \paragraph{Adding Lagrange multipliers.}
        To implement $F_\omega\ =0$, we introduce Lagrange multipliers represented by $\Lambda\in\scC^\infty(\Sigma,\frk)$, and we define the master action
        \begin{equation}\label{eq:MasterAction}
            S_\omega\ \coloneqq\ S+\tfrac12\int_\Sigma\inner{D(j_\omega)}{j_\omega}+\int_\Sigma\inner{\Lambda+D(g)g^{-1}}{F_\omega}~,
        \end{equation}
        where $S$ is any of the actions~\eqref{eq:PCMAction},~\eqref{eq:symmetricSpaceAction}, or~\eqref{eq:semiSymmetricSpaceAction} with $j$ replaced with $j_\omega$, and $D$ is the derivation introduced in~\eqref{eq:definitionD} by means of the two-cocycle $\Omega$. Without loss of generality, we have set the deformation parameter $\zeta$ equal to one. It should be noted that the action~\eqref{eq:MasterAction} is $\sfH$-gauge invariant in the case when considering coset superspaces $\sfG/\sfH$, that is, either~\eqref{eq:symmetricSpaceAction} or~\eqref{eq:semiSymmetricSpaceAction}, because of our assumption that $\Omega|_\frh$ vanishes; see \cref{sec:topologicalDeformations}. Moreover, in order to have the action~\eqref{eq:MasterAction} also $\sfK$-gauge invariant, we require that the gauge transformations~\eqref{eq:BGaugeTransformations} are augmented by
        \begin{equation}
            \Lambda\ \mapsto\ k^{-1}\Lambda k+k^{-1}D(k)
            \eforall
            k\ \in\ \scC^\infty(\Sigma,\sfK)~.
        \end{equation}

        Finally, we can simplify~\eqref{eq:MasterAction} as
        \begin{equation}\label{eq:MasterAction2}
            S_\omega\ =\ S+\tfrac12\int_\Sigma\inner{D(j_\omega)}{j_\omega}+\int_\Sigma\inner{\tilde\Lambda}{F_{j_\omega}}
            \ewith
            \tilde\Lambda\ \coloneqq\ g^{-1}\Lambda g+g^{-1}D(g)
        \end{equation}
        which is manifestly $\sfK$-gauge invariant. This form of the master action will be the starting point of our T-dualisation procedure.

        \paragraph{T-dualisation.}
        Upon integrating out $\Lambda$ in~\eqref{eq:MasterAction}, we recover the original sigma model action. Indeed, the equation of motion for $\Lambda$ is the flatness condition $F_\omega=0$, and hence, $\omega$ is locally pure gauge. Generally, however, $\omega$ might have non-trivial holonomies around non-contractible loops which prevents us from having $\omega$ be pure gauge globally. However, we shall ignore this technicality by assuming that the topology of $\Sigma$ is suitably chosen. Hence, we have $\omega$ pure gauge globally and so, we may fix a $\sfK$-gauge in which it vanishes identically.

        On the other hand, if we integrate out $\omega$, we obtain the T-dual model. Since the latter will still be invariant under $\sfK$-gauge transformation, we may fix a gauge to remove $\dim(\sfK)$ coordinates. For instance, for principal chiral models, this can be used to gauge away $\dim(\sfK)$ original coordinates, thus arriving at a gauge-fixed T-dual action with the same number of degrees of freedom as the original one, with the key difference that now the $\dim(\sfK)$ Lagrange multipliers play the role of dual coordinates. In particular, when taking $\sfK=\sfG$, gauge invariance can be used to set the original group element $g=\unit$, and, consequently, the T-dual model depends only on the dual coordinates $\Lambda$.

        In the case of sigma models defined on coset superspaces $\sfG/\sfH$, the reasoning is similar with the only difference being that the original model describes $\dim(\sfG)-\dim(\sfH)$ degrees of freedom. The gauge invariance inherited by the T-dual model has to be used to fix also some of the Lagrange multipliers~\cite{Sfetsos:1999zm,Lozano:2011kb}, while the remaining ones provide a set of dual coordinates.

        It should be stressed that this procedure can be complicated by certain uniqueness issues when trying to solve the equation of motion
        for $\omega$ for the projection of $j_\omega$ onto the Lie superalgebra of $\sfH$. While for the principal chiral model this issue does not arise, it may arise in symmetric and semi-symmetric models, and depends on the properties of the chosen Lie groups. In this case, the T-dualisation procedure needs to be modified accordingly and requires a case-by-case discussion. There are many examples in the literature for purely bosonic sigma models where this problem is automatically cured once the Lagrange multipliers are turned on in the master action~\eqref{eq:MasterAction} and the T-dualisation is performed before gauge fixing them~\cite{Sfetsos:1999zm,Lozano:2011kb,Borsato:2018idb}. Instead, for sigma models defined on supergroups the problem is made more severe by the appearance of fermionic coordinates, which are not invertible by their own nature. It turns out that in general switching on the fermionic Lagrange multipliers does not help in making the equations invertible, and more sophisticated approaches need to be used. In the case of purely fermionic cosets, this has been largely discussed in~\cite{Grassi:2011zf}. We shall come back to this in \cref{sec:cosets} and in a forthcoming paper.

        In what follows, we are going to discuss the T-dualisation of the three classes of models introduced above. Our primary focus is to identify explicitly the T-dual actions, study the associated Lax pair and the fate of integrability. In a simple example, corresponding to the principal chiral model describing an AdS$_3$ supergravity background, the question whether super non-Abelian T-duality leads to a dual background which still solves the supergravity torsion constraints is carefully investigated.

        \section{Super non-Abelian T-duality of principal chiral models}\label{sec:TDualityPCM}

        \subsection{General derivation of the T-dual model}

        We shall gauge a subsupergroup $\sfK$ of $\sfG$ so that the master action~\eqref{eq:MasterAction2} becomes
        \begin{equation}\label{eq:gaugedPCMAction}
            S_\omega\ =\ \tfrac12\int_\Sigma\inner{j_\omega}{\star j_\omega}+\tfrac12\int_\Sigma\inner{D(j_\omega)}{j_\omega}+\int_\Sigma\inner{\tilde\Lambda}{F_{j_\omega}}
        \end{equation}
        with $j_\omega$ as given in~\eqref{eq:gaugedFlatCurrent}.

        \paragraph{T-dual sigma model action.}
        The T-dual model is obtained by integrating out the $\omega$ field. In particular, upon varying~\eqref{eq:gaugedPCMAction} with respect to $\omega$, we obtain
        \begin{equation}\label{eq:EOM}
            \star j_\omega+\nabla_{j_\omega}\tilde\Lambda-D(j_\omega)\ =\ 0~.
        \end{equation}
        Upon making use of~\eqref{eq:ASDProjectors} and~\eqref{eq:definitionCovariantD}, this equation can be solved for $j_\omega$ as
        \begin{equation}\label{eq:solTdualPCMCurrent}
            j_\omega\ =\ -\frac{1}{1-D_{\tilde\Lambda}}\big(P^+(\rmd\tilde\Lambda)\big)+\frac{1}{1+D_{\tilde\Lambda}}\big(P^-(\rmd\tilde\Lambda)\big)\,.
        \end{equation}
        Therefore, with the help of the identity
        \begin{equation}\label{eq:variation2}
            \innerLarge{\frac{1}{1\pm D_U}(V)}{W}\ =\ \innerLarge{V}{\frac{1}{1\mp D_U}(W)}\eforall U,V,W\ \in\ \frg~,
        \end{equation}
        which is a consequence of~\eqref{eq:DInnerProductAction}, the T-dual action that follows from~\eqref{eq:gaugedPCMAction} upon substituting~\eqref{eq:solTdualPCMCurrent} is
        \begin{equation}\label{eq:PCMTdualAction}
            \tilde S\ =\ \int_\Sigma\innerLarge{\rmd\tilde\Lambda}{\frac{1}{1-D_{\tilde\Lambda}}\big(P^+(\rmd\tilde\Lambda)\big)}.
        \end{equation}

        \paragraph{T-dual equation of motion and Lax connection.}
        Using~\eqref{eq:variation2}, the expansion
        \begin{equation}\label{eq:variation1}
            \frac{1}{1\pm(D_{\tilde\Lambda}+\delta D_{\tilde\Lambda})}\ =\ \frac{1}{1\pm D_{\tilde\Lambda}}\mp\frac{1}{1\pm D_{\tilde\Lambda}}\circ\delta D_{\tilde\Lambda}\circ\frac{1}{1\pm D_{\tilde\Lambda}}+\cdots~,
        \end{equation}
        and $\delta D_{\tilde\Lambda}=\ad_{\delta\tilde\Lambda}$, some algebra shows that the equation of motion following from the T-dual action~\eqref{eq:PCMTdualAction} is
        \begin{equation}\label{eq:TdualEOM}
            \rmd\tilde j+\tfrac12[\tilde j,\tilde j]\ =\ 0~,
        \end{equation}
        where $\tilde j\coloneqq j_\omega$ with $j_\omega$ given in~\eqref{eq:solTdualPCMCurrent}. Consequently, $\tilde j$ satisfies~\eqref{eq:EOM}. Upon combining this with~\eqref{eq:TdualEOM} and making use of the Jacobi identity and the properties of $D$, we obtain
        \begin{equation}\label{eq:dualMC}
            \rmd{\star\tilde j}\ =\ 0~.
        \end{equation}
        The equations~\eqref{eq:TdualEOM} and~\eqref{eq:dualMC} confirm that for (deformed) principal chiral models the well-known pattern of exchanging equations of motions and Maurer--Cartan equations~\cite{Giveon:1993ai} occur also under super non-Abelian T-duality. Moreover, it allows to identify the T-dual Lax connection as
        \begin{equation}
            \tilde J(\tilde z)\ \coloneqq\ -\tfrac14(\tilde z-\tilde z^{-1})^2\,\tilde j+\tfrac14(\tilde z^2-\tilde z^{-2})\,{\star\tilde j}~,
        \end{equation}
        thus ensuring integrability of the T-dual model.

        \paragraph{T-dual Noether symmetries.}
        The T-dual model is invariant under global $\sfG$-transformations induced by the right action~\eqref{eq:rightGAction} and given by the transformation
        \begin{equation}\label{eq:deformedNoether}
            \tilde\Lambda\ \mapsto\ g_0^{-1}\tilde\Lambda g_0+g_0^{-1}D(g_0)
            \eforall
             g_0\ \in\ \sfG~,
        \end{equation}
        where $D(g_0)$ was defined in~\eqref{eq:extensionD}. Indeed, this follows immediately from~\eqref{eq:gaugeTransformationCovariantD}. The Noether current associated with this symmetry is then given by
        \begin{equation}\label{eq:TdualNoetherPCM}
            \tilde J_\rmN\ =\ {\star D_{\tilde\Lambda}(\tilde j)}~.
        \end{equation}

        \subsection{Example: \texorpdfstring{$\sfOSp(1|2)$}{OSp(1|2)} principal chiral model}\label{sec:OSp}

        As an explicit example of super non-Abelian T-duality, we shall now consider the principal chiral model for the orthosymplectic group $\sfOSp(1|2)$. The interest in this supergroup manifold is due to its interpretation as an $\mathcal N=1$ supersymmetric AdS$_3$, and as such, it represents an appropriate background for three-dimensional supergravity~\cite{Achucarro:1986uwr,Kuzenko:2011rd,Kuzenko:2012bc,Buchbinder:2017qls,Hutomo:2019mcx,Kuzenko:2021vmh}, as we are going to discuss.

        In the absence of supersymmetry, non-Abelian T-duality of the $\sfSL(2,\IR)$ principal chiral model, which describes AdS$_3$, has been performed in~\cite{Alvarez:1993qi}. There, it has been shown that the T-dualisation of an $\sfSL(2,\IR)$ subgroup of the isometry group $\sfSL(2,\IR)\times\sfSL(2,\IR)$ yields a three-dimensional metric corresponding to a black hole space-time and a non-trivial $B_2$-field. To the best of our knowledge, it is still unknown whether this black hole and $B_2$-field configuration is a solution of a matter-coupled gravity and whether it can be embedded into a three-dimensional supergravity.

        It is then interesting to investigate what happens in the case of the AdS$_3$ supergeometry. To this end, we shall discuss the T-dualisation of an $\sfOSp(1|2)$ subgroup of the group of superisometries $\sfOSp(1|2)\times\sfOSp(1|2)$. We will also comment about the T-dualisation of the maximal bosonic subgroup $\sfSL(2,\IR)\subseteq\sfOSp(1|2)$. We will find that in both cases, the target superspaces and $B_2$-fields of the T-dual models fall outside the class of possible solutions of a three-dimensional $\caN=1$ supergravity. This indicates that the black hole configuration obtained by the non-Abelian T-dualisation of the $\sfSL(2,\IR)$ principal chiral model in~\cite{Alvarez:1993qi} does not seem to have a conventional supergravity extension.

        \paragraph{Orthosymplectic algebra.}
        To begin with, we fix our conventions for the orthosymplectic algebra $\frosp(1|2)$. In spinorial notation, it is described by a set of two fermionic generators $Q_{\alpha}$ and three bosonic generators $L_{\alpha\beta}=L_{\beta\alpha}$ for $\alpha,\beta,\ldots=1,2$ subject to the following non-trivial commutation relations
        \begin{equation}\label{eq:osp12CommutationRelations}
            \begin{gathered}
                [L_{\alpha\beta},L_{\gamma\delta}]\ =\ -\rmi\big(\eps_{\gamma(\alpha}L_{\beta)\delta}+\eps_{\delta(\alpha}L_{\beta)\gamma}\big)\,,
                \\
                [Q_\alpha, Q_\beta]\ =\ L_{\alpha\beta}~,
                \quad
                [L_{\alpha\beta},Q_\gamma]\ =\ -\rmi\eps_{\gamma(\alpha}Q_{\beta)}~.
            \end{gathered}
        \end{equation}
        The parentheses on the right-hand sides denote normalised symmetrisation of the enclosed indices and $\rmi$ is the imaginary unit. In addition, $\eps_{\alpha\beta}=-\eps_{\beta\alpha}$ with $\eps_{\alpha\gamma}\eps^{\gamma\beta}=\delta_\alpha{}^\beta$ and $\delta_\alpha{}^\beta$ is the Kronecker symbol. Furthermore, we introduce an inner product $\inner{-}{-}$ on $\frosp(1|2)$ by
        \begin{equation}\label{eq:innerProduct}
            \inner{Q_\alpha}{Q_\beta}\ \coloneqq\ \rmi\eps_{\alpha\beta}~,
            \quad
            \inner{L_{\alpha\beta}}{L_{\gamma\delta}}\ \coloneqq\ \eps_{\alpha(\gamma}\eps_{\delta)\beta}~,
            \eand
            \inner{Q_\alpha}{L_{\beta\gamma}}\ \coloneqq\ 0~.
        \end{equation}

        \paragraph{Sigma model action.}
        Since $H^2(\frosp(1|2))$ vanishes~\cite{Scheunert:1997ks}, we work with the undeformed action~\eqref{eq:PCMAction}. In particular, we parametrise $g\in\sfG$ as
        \begin{equation}\label{eq:osp-parametrisation}
            g\ \coloneqq\ g_\ttb g_\ttf
            \ewith
            g_\ttb\ \coloneqq\ \rme^{x^{\alpha\beta}L_{\alpha\beta}}~,~~
            g_\ttf\ \coloneqq\ \rme^{-\caF}~,
            \eand
            \caF\ \coloneqq\ \theta^\alpha Q_\alpha~,
        \end{equation}
        with three bosonic coordinates $x^{\alpha\beta}=x^{\beta\alpha}$ and two fermionic coordinates $\theta^\alpha$. The corresponding current $j=g^{-1}\rmd g\eqqcolon j^{\alpha\beta}L_{\alpha\beta}+j^\alpha Q_\alpha$ can be split as
        \begin{subequations}
            \begin{equation}
                j\ =\ j_\ttb+j_\ttf~,
            \end{equation}
            where (see also \cref{app:fermionicCurrent})
            \begin{equation}\label{eq:fermionicCurrent}
                j_\ttb\ \coloneqq\ g_\ttb^{-1}\rmd g_\ttb
                \eand
                j_\ttf\ \coloneqq\ -\nabla_{j_\ttb}\caF-\tfrac12\ad_\caF(\nabla_{j_\ttb}\caF)-\tfrac16\ad_\caF^2(\rmd\caF)
            \end{equation}
        \end{subequations}
        and the covariant derivative is as defined in~\eqref{eq:covariantDerivative}. Explicitly,
        \begin{equation}
            \begin{aligned}
                j_\ttb\ &\eqqcolon\ e^{\alpha\beta}L_{\alpha\beta}~,
                \\
                j_\ttf\ &=\ -\big[\big(1-\tfrac\rmi8\theta^2\big)\rmd\theta^\alpha+\rmi\theta_\alpha e^{\alpha\beta}\big]Q_\beta+\tfrac12(\theta^\alpha\rmd\theta^\beta+\tfrac\rmi2\theta^2 e^{\alpha\beta})L_{\alpha\beta}~,
            \end{aligned}
        \end{equation}
        where $\theta^2\coloneqq\theta^\alpha\theta_\alpha$. Consequently,
        \begin{equation}\label{eq:josp}
            \begin{aligned}
                j\ &=\ -\big(1-\tfrac\rmi8\theta^2\big)\big(\rmd\theta^\beta+\rmi\theta_\alpha e^{\alpha\beta}\big)Q_\beta+\big(1+\tfrac\rmi4\theta^2\big)\big(e^{\alpha\beta}+\tfrac12\theta^{(\alpha}\rmd\theta^{\beta)}\big)L_{\alpha\beta}~.
            \end{aligned}
        \end{equation}
        In order to compute $e^{\alpha\beta}$ explicitly, we use the Baker--Campbell--Hausdorff formula together with the identities~\eqref{eq:XYbosonicCommutator} specialised to $U\equiv x^{\alpha\beta}L_{\alpha\beta}$ and $V\equiv\rmd U$. Indeed, $j_\ttb=\rme^{-U}\rmd\rme^U=\sum_{k=0}^\infty\frac{(-1)^k}{(k+1)!}\,\ad^k_{U}(\rmd U)$, and so
        \begin{subequations}
            \begin{equation}\label{eq:OSpjbos}
                \begin{aligned}
                    e^{\alpha\beta}\ &=\ \rmd x^{\alpha\beta}-\frac{2x-\sqrt{2}\sinh(\sqrt{2}x)}{4x^3}(x^2\rmd x^{\alpha\beta}-2x_\gamma{}^\alpha x_\delta{}^\beta\rmd x^{\gamma\delta})+\frac{2\rmi\sinh^2\big(\frac{x}{\sqrt{2}}\big)}{x^2}x_\gamma{}^{(\alpha}\rmd x^{\beta)\gamma}~.
                \end{aligned}
            \end{equation}
            where
            \begin{equation}\label{eq:DefOfx2}
                {x_\gamma}^\alpha{x_\beta}^\gamma\ =\ -\tfrac12{\delta_\beta}^\alpha x_{\gamma\delta}x^{\gamma\delta}\ \eqqcolon\ -\tfrac12{\delta_\beta}^\alpha x^2
                \quad\Rightarrow\quad
                x^2\ =\ x_{\alpha\beta}x^{\alpha\beta}~.
            \end{equation}
        \end{subequations}
        Note that the series in~\eqref{eq:OSpjbos} are in powers of $x^2$ only.

        Upon inserting these expressions into~\eqref{eq:PCMAction} and using~\eqref{eq:innerProduct}, we eventually find
        \begin{equation}\label{eq:osp-initial-action}
            \begin{aligned}
                S\ &=\ -\tfrac12\int_\Sigma\Big\{j^{\alpha\beta}\wedge{\star j_{\alpha\beta}}+\rmi j^\alpha\wedge{\star j_\alpha}\Big\}
                \\
                &=\ -\tfrac12\int_\Sigma\Big\{e^{\alpha\beta}\wedge{\star e_{\alpha\beta}}-e^{\alpha\beta}\wedge{\star\theta_{(\alpha}\rmd\theta_{\beta)}}+\rmi\big(1-\tfrac{\rmi}{16}\theta^2\big)\rmd\theta^\alpha\wedge{\star\rmd\theta_\alpha}\Big\}\,.
            \end{aligned}
        \end{equation}
        Furthermore, in the metric-like form, the action reads as
        \begin{subequations}
            \begin{equation}
                S\ =\ \int_\Sigma\Big\{\tfrac12\rmd x^{\gamma\delta}\wedge{\star \rmd x^{\alpha\beta}}g_{\alpha\beta,\gamma\delta}+\rmd x^{\beta\gamma}{\wedge{\star\rmd\theta^\alpha}}g_{\alpha,\beta\gamma}+\tfrac12\rmd\theta^{\beta}\wedge{\star\rmd\theta^\alpha}g_{\alpha,\beta}\Big\}
            \end{equation}
            with the components
            \begin{equation}
                \begin{gathered}
                    g_{\alpha\beta,\gamma\delta}\ \coloneqq\ g_{\gamma\delta,\alpha\beta}\ \coloneqq\ g_1\eps_{\alpha(\gamma}\eps_{\delta)\beta}+g_2x_{\alpha(\gamma}x_{\delta) \beta}~,
                    \\
                    g_{\alpha,\beta\gamma}\ \coloneqq\ g_{\beta\gamma,\alpha}\ \coloneqq\ \big[g_3\eps_{\alpha(\beta}\eps_{\gamma)\delta}+g_4x_{\alpha(\beta}x_{\gamma)\delta} +g_5(x_{\alpha(\beta}\eps_{\gamma)\delta}+x_{\delta(\beta}\eps_{\gamma)\alpha})\big]\theta^\delta~,
                    \\
                    g_{\alpha,\beta}\ \coloneqq\ -g_{\beta,\alpha}\ \coloneqq\ \eps_{\alpha\beta}\bigg(\rmi+\frac{\theta^2}{16}\bigg)
                \end{gathered}
            \end{equation}
            and the coefficient functions
            \begin{equation}
                \begin{gathered}
                    g_1\ \coloneqq\ -\frac{1-x^2-\cosh{(\sqrt{2}x)}}{2x^2}~,
                    \quad
                    g_2\ \coloneqq\ -\frac{1+x^2-\cosh{(\sqrt{2}x)}}{x^4}~,
                    \\
                    g_3\ \coloneqq\ \frac{2x+\sqrt{2}\sinh{(\sqrt{2}x)}}{8x}~,
                    \quad
                    g_4\ \coloneqq\ \frac{-2x+\sqrt{2}\sinh{(\sqrt{2}x)}}{4x^3}~,
                    \quad
                    g_5\ \coloneqq\ \frac{\rmi}{2x^2}\sinh^2\bigg(\frac{x}{\sqrt{2}}\bigg)\,.
                \end{gathered}
            \end{equation}
        \end{subequations}

        \paragraph{T-dual sigma model action.}
        We now perform super non-Abelian T-duality along $\sfOSp(1|2)$ corresponding to the left action~\eqref{eq:leftGAction}. The general expression for the T-dual action of the undeformed principal chiral model is given in~\eqref{eq:PCMTdualAction} with $D_{\tilde\Lambda}$ replaced with $\ad_{\tilde\Lambda}$. Moreover, since we are gauging the whole group, we can fix a gauge in which $g=\unit$. Therefore, the T-dual action becomes
        \begin{equation}\label{eq:PCMTdualActionOsp}
            \tilde S\ =\ \tfrac12\int_\Sigma\innerLarge{\rmd\Lambda}{\frac{1}{1-\ad_{\Lambda}}(\rmd\Lambda+\star\rmd\Lambda)}.
        \end{equation}
        In order to compute it explicitly, we rearrange the integrand as outlined in \cref{app:Liealgebraidentities}. In particular, upon specialising the identities~\eqref{eq:finalexpansion} to $U\equiv\Lambda$ and $V\equiv\rmd\Lambda$ with $\Lambda$ expanded in terms of a set of dual coordinates $\tilde x^{\alpha\beta}$ and $\tilde\theta^\alpha$ as $\Lambda\eqqcolon\tilde x^{\alpha\beta}L_{\alpha\beta}+\tilde\theta^\alpha Q_\alpha$, we obtain
        \begin{subequations}\label{eq:TDualVielBeinsOSP}
            \begin{equation}
                \begin{aligned}
                    \frac{1}{1-\ad_\Lambda}(\rmd\Lambda)\ &=\ \frac{1}{1-2\tilde x^2}\big[(1-\tilde x^2)Z^{\alpha\beta}-2\tilde x_\gamma{}^\alpha\tilde x_\delta{}^\beta Z^{\gamma\delta}-2\rmi\tilde x_\gamma{}^{(\alpha}Z^{\beta)\gamma}\big]L_{\alpha\beta}
                    \\
                    &\kern1cm+\frac{2}{2-\tilde x^2}\zeta^\alpha({\delta_\alpha}^\beta-\rmi\tilde x_\alpha{}^\beta)Q_\beta~,
                \end{aligned}
            \end{equation}
            where
            \begin{equation}
                \begin{aligned}
                    Z^{\alpha\beta}\ &=\ \rmd\tilde x^{\alpha\beta}+\frac{2}{2-\tilde x^2}\big[\rmd\tilde\theta^{(\alpha}-\rmi\rmd\tilde\theta^\gamma\tilde x_\gamma{}^{(\alpha}\big]\tilde\theta^{\beta)}
                    \\
                    &\kern1cm+\frac{\rmi}{(2-\tilde x^2)(1-2\tilde x^2)}\Big[\big(1-\tfrac12\tilde x^2\big)\rmd\tilde x^{\alpha\beta}-3\tilde x_\gamma{}^\alpha\tilde x_\delta{}^\beta\rmd\tilde x^{\gamma\delta}-3\rmi\tilde x_\gamma{}^{(\alpha}\rmd\tilde x^{\beta)\gamma}\Big]\tilde\theta^2~,
                    \\
                    \zeta^\alpha\ &=\ \rmd\tilde\theta^\alpha-\frac{\rmi}{1-2\tilde x^2}\big[(1-\tilde x^2)\rmd\tilde x^{\alpha\beta}-2\tilde x_\gamma{}^\alpha\tilde x_\delta{}^\beta\rmd\tilde x^{\gamma\delta}-2\rmi\tilde x_\gamma{}^{(\alpha}\rmd\tilde x^{\beta)\gamma}\big]\tilde\theta_\beta
                    \\
                    &\kern1cm-\frac{\rmi}{(2-\tilde x^2)(1-2\tilde x^2)}\Big[\tfrac32\rmd\tilde\theta^\alpha-\rmi\big(\tfrac72-\tilde x^2\big)\rmd\tilde\theta^\beta\tilde x_\beta{}^\alpha\Big]\tilde\theta^2
                \end{aligned}
            \end{equation}
        \end{subequations}
        and $\tilde x^2$ is as defined in~\eqref{eq:DefOfx2}. Therefore, summing everything and using the inner product~\eqref{eq:innerProduct}, the explicit expression for the T-dual action~\eqref{eq:PCMTdualActionOsp} eventually reads
        \begin{equation}
            \begin{aligned}
                \tilde S\ &=\ -\frac12\int_\Sigma\left\{\frac{\rmd\tilde x_{\alpha\beta}\wedge\star\big[(1-\tilde x^2)Z^{\alpha\beta}-2\tilde x_\gamma{}^\alpha\tilde x_\delta{}^\beta Z^{\gamma\delta}-2\rmi\tilde x_\gamma{}^{(\alpha}Z^{\beta)\gamma}\big]}{1-2\tilde x^2}\right.
                \\
                &\kern2.5cm+\frac{\rmd\tilde x_{\alpha\beta}\wedge\big[(1-\tilde x^2)Z^{\alpha\beta}-2\tilde x_\gamma{}^\alpha\tilde x_\delta{}^\beta Z^{\gamma\delta}-2\rmi\tilde x_\gamma{}^{(\alpha}Z^{\beta)\gamma}\big]}{1-2\tilde x^2}
                \\
                &\kern3cm\left.+\,\frac{2\rmi\rmd\tilde\theta^\alpha\wedge{\star\zeta^\beta(\eps_{\alpha\beta}-\rmi\tilde x_{\alpha\beta})}}{2-\tilde x^2}+\frac{2\rmi\rmd\tilde\theta^\alpha\wedge\zeta^\beta(\eps_{\alpha\beta}-\rmi\tilde x_{\alpha\beta})}{2-\tilde x^2}\right\}.
            \end{aligned}
        \end{equation}
        In the metric-like form, the action takes the following form in which one can immediately notice the presence of a two-form superfield $\tilde B_2$ which was absent in the original $\sfOSp(1|2)$ model,
        \begin{subequations}\label{eq:TdualOSpMetricForm}
            \begin{equation}
                \begin{aligned}
                    S\ &=\ \int_\Sigma\Big\{\tfrac12\rmd\tilde x^{\gamma\delta}\wedge{\star\rmd\tilde x^{\alpha\beta}}\tilde g_{\alpha\beta,\gamma\delta}+\rmd\tilde x^{\beta\gamma}{\wedge{\star\rmd\tilde\theta^\alpha}}\tilde g_{\alpha,\beta\gamma}+\tfrac12\rmd\tilde\theta^{\beta}\wedge{\star\rmd\theta^\alpha}\tilde g_{\alpha,\beta}
                    \\
                    &\kern2cm+\tfrac12\rmd\tilde x^{\gamma\delta}\wedge\rmd\tilde x^{\alpha\beta}\tilde B_{\alpha\beta,\gamma\delta}+\rmd\tilde x^{\beta\gamma}\wedge\rmd\tilde\theta^\alpha\tilde B_{\alpha,\beta\gamma}+\tfrac12\rmd\tilde\theta^{\beta}\wedge\rmd\theta^\alpha\tilde B_{\alpha,\beta}\Big\}
                \end{aligned}
            \end{equation}
            with the components
            \begin{equation}
                \begin{gathered}
                    \tilde g_{\alpha\beta,\gamma\delta}\ \coloneqq\ \tilde g_{\gamma\delta,\alpha\beta}\ \coloneqq\ \big(\tilde g_1+\tilde g_3\tilde\theta^2\big)\eps_{\alpha(\gamma}\eps_{\delta)\beta}+\big(\tilde g_2+\tilde g_4\tilde\theta^2\big)\tilde{x}_{\alpha(\gamma}\tilde{x}_{\delta) \beta}~,
                    \\
                    \tilde g_{\alpha,\beta\gamma}\ \coloneqq\ \tilde g_{\beta\gamma,\alpha}\ \coloneqq\ \big(\tilde g_5\tilde{x}_{\alpha(\beta}\eps_{\gamma)\delta}+\tilde g_6\tilde{x}_{\delta(\beta}\eps_{\gamma)\alpha}\big)\tilde\theta^\delta~,
                    \\
                    \tilde g_{\alpha,\beta}\ \coloneqq\ -\tilde g_{\beta,\alpha}\ \coloneqq\ \eps_{\alpha\beta}\big(\tilde g_7+\tilde g_8\tilde\theta^2\big)~,
                    \\[5pt]
                    \tilde B_{\alpha\beta,\gamma\delta}\ \coloneqq\ -\tilde B_{\gamma\delta,\alpha\beta}\ \coloneqq\ \big(\tilde b_1+\tilde b_2\tilde\theta^2\big)(\tilde x_{\alpha(\gamma}\eps_{\delta)\beta}+\tilde x_{\beta(\gamma}\eps_{\delta)\alpha})~,
                    \\
                    \tilde B_{\alpha,\beta\gamma}\ \coloneqq\ -\tilde B_{\beta\gamma,\alpha}\ \coloneqq\ \big(\tilde b_3\eps_{\alpha(\beta}\eps_{\gamma)\delta}+\tilde b_4\tilde x_{\alpha(\beta}\tilde x_{\gamma)\delta}\big)\tilde\theta^\delta~,
                    \\
                    \tilde B_{\alpha,\beta}\ \coloneqq\ \tilde B_{\beta,\alpha}\ \coloneqq\ \tilde x_{\alpha\beta}\big(\tilde b_5+\tilde b_6\tilde\theta^2\big)
                \end{gathered}
            \end{equation}
            and the coefficient functions
            \begin{equation}
                \begin{gathered}
                    \tilde g_1\ \coloneqq\ \frac{1-\tilde x^2}{1-2\tilde x^2}~,
                    \quad
                    \tilde g_2\ \coloneqq\ \frac{2}{1-2\tilde x^2}~,
                    \quad
                    \tilde g_3\ \coloneqq\ \frac{\rmi(1+2\tilde x^4)}{(2-\tilde x^2)(1-2\tilde x^2)^2}~,
                    \quad
                    \tilde g_4\ \coloneqq\ \frac{4\rmi}{(1-2\tilde x^2)^2}~,
                    \\
                    \tilde g_5\ \coloneqq\ \frac{2\rmi}{1-2\tilde x^2}~,
                    \quad
                    \tilde g_6\ \coloneqq\ \frac{2\rmi(1+\tilde x^2)}{(2-\tilde x^2)(1-2\tilde x^2)}~,
                    \quad
                    \tilde g_7\ \coloneqq\ \frac{2\rmi}{2-\tilde x^2}~,
                    \\
                    \tilde g_8\ \coloneqq\ \frac{(6+7\tilde x^2 -2\tilde x^4)}{2(2-\tilde x^2)^2(1-2\tilde x^2)}~,
                    \\[5pt]
                    \tilde b_1\ \coloneqq\ \frac{\rmi}{1-2\tilde x^2}~,
                    \quad
                    \tilde b_2\ \coloneqq\ -\frac{5+2\tilde x^2}{2(2-\tilde x^2)(1-2\tilde x^2)^2}~,
                    \\
                    \tilde b_3\ \coloneqq\ \frac{1}{1-2\tilde x^2}~,
                    \quad
                    \tilde b_4\ \coloneqq\ \frac{6}{(2-\tilde x^2)(1-2\tilde x^2)}~,
                    \\
                    \tilde b_5\ \coloneqq\ -\frac{2}{2-\tilde x^2}~,
                    \quad
                    \tilde b_6\ \coloneqq\ \frac{\rmi(10-2\tilde x^2)}{(2-\tilde x^2)^2(1-2\tilde x^2)}~,
                \end{gathered}
            \end{equation}
        \end{subequations}
        and $\tilde x^2$ is as defined in~\eqref{eq:DefOfx2}. Note that the remaining isometry group is $\sfOSp(1|2)$ as follows from the general discussion around~\eqref{eq:TdualNoetherPCM}. Hence, half of the supersymmetries of the original model are broken. As shown in \cref{app:SL2-dualisation}, this also happens when dualising only the maximal bosonic subgroup $\sfSL(2,\IR)$. This is in agreement with what was already established, for instance, in the cases of purely bosonic non-Abelian T-duality along the AdS$_3$- and $S^3$-directions in AdS$_3\times S^3\times{\rm CY}_2$ backgrounds~\cite{Sfetsos:2010uq,Lozano:2011kb,Ramirez:2021tkd} (where ${\rm CY}_2$ is a Calabi--Yau two-fold).

        \paragraph{T-dual supervielbeins.}
        We observe that the dual action~\eqref{eq:PCMTdualActionOsp} can be rewritten as
        \begin{equation}\label{eq:PCMTdualActionInVielBeins}
            \tilde S\ =\ \tfrac12\int_\Sigma\big\{\inner{\tilde\bme}{\star\tilde\bme}+\inner{\tilde\bme}{\ad_\Lambda(\tilde\bme)}\big\}
            \ewith
            \tilde\bme\ \coloneqq\ -\frac{1}{1-\ad_\Lambda}(\rmd\Lambda)~.
        \end{equation}
        Consequently, $\tilde\bme$ represents the T-dual supervielbein.\footnote{Note that $\frac{1}{1+\ad_{\Lambda}}(\rmd\Lambda)$ is an equivalent form for the supervielbeins as follows from the identity~\eqref{eq:variation2}.} A short calculation reveals that
        \begin{equation}\label{eq:MCTdualVielBeinPCM}
            \rmd\tilde\bme\ =\ -\frac12[\tilde\bme,\tilde\bme]-\frac12\frac{1}{1-\ad_{\Lambda}}([\tilde\bme,\tilde\bme])~.
        \end{equation}
        Furthermore, from the action~\eqref{eq:PCMTdualActionInVielBeins} we identify the $\tilde B_2$-field as
        \begin{equation}
            \tilde B_2\ \coloneqq\ \tfrac12\inner{\tilde\bme}{\ad_\Lambda(\tilde\bme)}~.
        \end{equation}
        A short calculation then leads to
        \begin{equation}\label{eq:dualPCMH3}
            \tilde H_3\ \coloneqq\ \rmd\tilde B_2\ =\ \frac12\innerLarge{\tilde\bme}{\frac{1}{1-\ad_\Lambda}([\tilde\bme,\tilde\bme])}.
        \end{equation}
        To apply these formu{\ae} to the $\sfOSp(1|2)$ case, we consider the expansions $\tilde\bme\eqqcolon\tilde\bme^{\alpha\beta}L_{\alpha\beta}+\tilde\bme^\alpha Q_\alpha$ and $\Lambda\eqqcolon \tilde x^{\alpha\beta}L_{\alpha\beta}+\tilde\theta^\alpha Q_\alpha$, and compute
        \begin{equation}\label{eq:ee}
            [\tilde\bme,\tilde\bme]\ =\ (2\rmi\tilde\bme^{\gamma(\alpha}\wedge\tilde\bme_\gamma{}^{\beta)}-\tilde\bme^\alpha\wedge\tilde\bme^\beta)L_{\alpha\beta}+2\rmi\tilde\bme^{\alpha\beta}\wedge\tilde\bme_\beta Q_\alpha\ \eqqcolon\ \upsilon^{\alpha\beta}L_{\alpha\beta}+\upsilon^\alpha Q_\alpha~,
        \end{equation}
        the equations in~\eqref{eq:TDualVielBeinsOSP} imply that
        \begin{subequations}
            \begin{equation}\label{eq:someAdAction}
                \begin{aligned}
                    \frac{1}{1-\ad_\Lambda}([\tilde\bme,\tilde\bme])\ &=\ \frac{1}{1-2\tilde x^2}\big[(1-\tilde x^2)Z^{\alpha\beta}-2\tilde x_\gamma{}^\alpha\tilde x_\delta{}^\beta Z^{\gamma\delta}-2\rmi\tilde x_\gamma{}^{(\alpha}Z^{\beta)\gamma}\big]L_{\alpha\beta}
                    \\
                    &\kern1cm+\frac{2}{2-\tilde x^2}\zeta^\alpha({\delta_\alpha}^\beta-\rmi\tilde x_\alpha{}^\beta)Q_\beta
                \end{aligned}
            \end{equation}
            with
            \begin{equation}
                \begin{aligned}
                    Z^{\alpha\beta}\ &=\ \upsilon^{\alpha\beta}+\frac{2}{2-\tilde x^2}\big[\upsilon^{(\alpha}-\rmi\upsilon^\gamma\tilde x_\gamma{}^{(\alpha}\big]\tilde\theta^{\beta)}
                    \\
                    &\kern1cm+\frac{\rmi}{(2-\tilde x^2)(1-2\tilde x^2)}\Big[\big(1-\tfrac12\tilde x^2\big)\upsilon^{\alpha\beta}-3\tilde x_\gamma{}^\alpha\tilde x_\delta{}^\beta\upsilon^{\gamma\delta}-3\rmi\tilde x_\gamma{}^{(\alpha}\upsilon^{\beta)\gamma}\Big]\tilde\theta^2~,
                    \\
                    \zeta^\alpha\ &=\ \upsilon^\alpha-\frac{\rmi}{1-2\tilde x^2}\big[(1-\tilde x^2)\upsilon^{\alpha\beta}-2\tilde x_\gamma{}^\alpha\tilde x_\delta{}^\beta\upsilon^{\gamma\delta}-2\rmi\tilde x_\gamma{}^{(\alpha}\upsilon^{\beta)\gamma}\big]\tilde\theta_\beta
                    \\
                    &\kern1cm-\frac{\rmi}{(2-\tilde x^2)(1-2\tilde x^2)}\Big[\tfrac32\upsilon^\alpha-\rmi\big(\tfrac72-\tilde x^2\big)\upsilon^\beta\tilde x_\beta{}^\alpha\Big]\tilde\theta^2~.
                \end{aligned}
            \end{equation}
        \end{subequations}
        Therefore,~\eqref{eq:MCTdualVielBeinPCM} becomes
        \begin{subequations}
            \begin{equation}\label{eq:Lprojection}
                \rmd\tilde\bme^{\alpha\beta}\ =\ -\frac12\upsilon^{\alpha\beta}-\frac{1}{2(1-2\tilde x^2)}\big[(1-\tilde x^2)Z^{\alpha\beta}-2\tilde x_\gamma{}^\alpha\tilde x_\delta{}^\beta Z^{\gamma\delta}-2\rmi\tilde x_\gamma{}^\alpha Z^{\beta\gamma}\big]
            \end{equation}
            and
            \begin{equation}
                \rmd\tilde\bme^\alpha\ =\ -\frac12\upsilon^\alpha-\frac{2}{2(2-\tilde x^2)}\zeta^\alpha({\delta_\alpha}^\beta-\rmi\tilde x_\alpha{}^\beta)~.
            \end{equation}
        \end{subequations}

        \paragraph{$\sfSL(2,\IR)$ T-dual model.}
        Note that upon setting $\tilde\theta^\alpha=0$, the model~\eqref{eq:TdualOSpMetricForm} reduces to the T-dual model found in~\cite{Alvarez:1993qi} when dualising the $\sfSL(2,\IR)$ principal chiral model. In particular, the target space metric $\tilde g_{\alpha\beta\,,\gamma\delta}$ for $\tilde\theta^\alpha=0$ in~\eqref{eq:TdualOSpMetricForm} was interpreted in~\cite{Alvarez:1993qi} as a black hole that is asymptotically anti-de Sitter with the scalar curvature
        \begin{equation}\label{eq:dualRBH}
            \tilde R\ =\ -2\frac{4\tilde x^4-6\tilde x^2+9}{(1-2\tilde x^2)^2}
        \end{equation}
        for the Levi-Civita connection. Furthermore, we observe that the $\tilde B_2$-field $\tilde B_{\alpha\beta\,,\gamma\delta}$ for $\tilde\theta^\alpha=0$ in~\eqref{eq:TdualOSpMetricForm} has a non-constant field strength
        \begin{equation}\label{eq:dualSL2H3}
            \tilde H_3\ =\ \frac\rmi3\frac{3-2\tilde x^2}{1-2\tilde x^2}\,\tilde\bme_\alpha{}^\beta\wedge\tilde\bme_\beta{}^\gamma\wedge\tilde\bme_\gamma{}^\alpha~,
        \end{equation}
        as follows from~\eqref{eq:dualPCMH3} together with~\eqref{eq:someAdAction} after some algebra. Being non-constant, the curvature~\eqref{eq:dualRBH} and the field strength~\eqref{eq:dualSL2H3} cannot be a solution of Einstein gravity coupled only to a $B_2$-field. Indeed, the equations of motion for this theory are
        \begin{equation}
            \begin{aligned}
                R_{\mu\nu}-\tfrac12g_{\mu\nu}R\ &=\ -\tfrac14\big(H_{\mu\kappa\lambda}H_{\nu}{}^{\kappa\lambda}-\tfrac 16g_{\mu\nu}H_{\kappa\lambda\sigma}H^{\kappa\lambda\sigma}\big)\,,
                \\
                \nabla^\mu H_{\mu\nu\kappa}\ &=\ 0~,
            \end{aligned}
        \end{equation}
        where $R_{\mu\nu}$ is the Ricci tensor for the Levi-Civita connection of the metric $g_{\mu\nu}$, $\nabla_\mu$ the covariant derivative with respect to the Levi-Civita connection, and $H_{\mu\nu\kappa}$ the components of $H_3$. The second equation implies that $H_{\mu\nu\kappa}=c\,\sqrt{|g|}\,\eps_{\mu\nu\kappa}$ with $\eps_{\mu\nu\kappa}$ the Levi-Civita symbol and $c\in\IR$ some constant. Thus, $H_3$ must be constant. Moreover, the trace over the first equation yields that $R=\frac14H_{\mu\nu\kappa}H^{\mu\nu\kappa}=\frac32c^2$ and so, also the scalar curvature must be constant. Since neither the scalar curvature~\eqref{eq:dualRBH} is constant nor the three-form field strength~\eqref{eq:dualSL2H3} is of the desired form, we conclude that the bosonic T-dual model is not a solution of Einstein gravity coupled only to a $B_2$-field. Therefore, the metric $\tilde g_{\alpha\beta\,,\gamma\delta}$ and the $\tilde B_2$-field $\tilde B_{\alpha\beta\,,\gamma\delta}$ may only be solutions in a theory in which gravity and the $B_2$-field couple to some other matter fields.\footnote{Note that the T-dualisation also yields a non-trivial dilaton $\tilde\phi=-\log(1-2\tilde x^2)$ but the standard coupling of the dilaton to Einstein gravity and a $B_2$-field does not solve this issue for the scalar curvature~\eqref{eq:dualRBH} and three-form field strength~\eqref{eq:dualSL2H3}.} It would be of interest to find this theory.

        \paragraph{Supergravity constraints for the $\sfOSp(1|2)$ T-dual model.}
        We will not elaborate on the above issue of the bosonic $\sfSL(2,\IR)$ model  here but instead ask the question whether the three-dimensional black hole of~\cite{Alvarez:1993qi} can be part of a conventional three-dimensional supergravity. In other words, we shall study whether the target superspace of the super non-Abelian T-dual model obtained from the $\sfOSp(1|2)$ principal chiral model may be interpreted as a supersymmetric black hole which might be a solution of a three-dimensional $\caN=1$ supergravity. For this to be possible, the target superspace of the T-dual model should respect an additional key feature: it should be understood as an appropriate supergravity background. Put differently, it should satisfy the superspace supergravity constraints, which in three-dimensions are off-shell constraints that do not imply the on-shell supergravity equations of motion.

        A brief review of three-dimensional  supergravity is given in \cref{app:conformalSUGRA}. In particular, we are interested in the torsion constraints~\eqref{eq:3dConformalSUGRATorsion} and the constraint~\eqref{eq:H3kuzenko} on the three-form field strength $H_3$ of the $B_2$-field.

        The target superspace of the original $\sfOSp(1|2)$ principal chiral model is an AdS$_3$ solution of three-dimensional $\caN=1$ supergravity. As such, it satisfies the constraints~\eqref{eq:3dConformalSUGRATorsion}. Indeed, the supervielbeins can be read off the flat current $j=g^{-1}\rmd g$ given in~\eqref{eq:josp}. In particular, expanding $j\eqqcolon j^{\alpha\beta}L_{\alpha\beta}+j^\alpha Q_\alpha$ and using the algebra~\eqref{eq:osp12CommutationRelations}, the flatness equation~\eqref{eq:MCEquation} in components reads
        \begin{equation}
            \rmd j^{\alpha\beta}+\rmi\,j^{\gamma\alpha}\wedge j_\gamma{}^{\beta}\ =\ \tfrac12 j^\alpha\wedge j^\beta
            \eand
            \rmd j^\alpha+\rmi\,j^\beta\wedge j_\beta{}^\alpha\ =\ 0~.
        \end{equation}
        It follows that the torsion constraints~\eqref{eq:3dConformalSUGRATorsion} are satisfied, provided that we identify the supervielbeins, the connection one-form, and the non-vanishing components of the torsion as
        \begin{equation}
            \bme^{\alpha\beta}\ =\ -j^{\alpha\beta}~,
            \quad
            \bme^\alpha\ =\ -j^\alpha~,
            \quad
            \bmomega_\alpha{}^\beta\ =\ -\tfrac{\rmi}{2}\,j_\alpha{}^\beta~,
            \eand
            \bmT_{\alpha\beta\,\gamma}{}^\delta\ =\ -\,\eps_{\gamma(\alpha}\delta_{\beta)}{}^\delta~.
        \end{equation}
        Since $j_{\alpha\beta}=j_{\beta\alpha}$, the metric compatibility~\eqref{eq:3dMetricCompatibility} is also satisfied. Though the $\sfOSp(1|2)$ principal chiral model does not include the $B_2$-field, its field strength $H_3$ can be a dynamical source of the cosmological constant in the supergravity equations of motion ensuring the existence of the AdS$_3$ superspace solution (see e.g.~\cite{Buchbinder:2017qls}). For the AdS$_3$ solution,
        \begin{equation}\label{eq:H3ads3}
            H_3\ =\ \bme_\alpha\wedge\bme^{\alpha\beta}\wedge\bme_\beta+\tfrac{4\rmi}{3\ell}\bme_\alpha{}^\beta\wedge\bme_\beta{}^\gamma\wedge\bme_\gamma{}^\alpha\,,
        \end{equation}
        where $\ell$ is proportional to the AdS$_3$ radius. This $H_3$ satisfies the constraint~\eqref{eq:H3kuzenko}.

        Now we would like to check whether the dual supergeometry satisfies the supergravity constraints. Let us focus on the first torsion constraint in~\eqref{eq:3dConformalSUGRATorsion}, that is, the projection~\eqref{eq:Lprojection} of~\eqref{eq:MCTdualVielBeinPCM} onto $L_{\alpha\beta}$. The torsion constraint is satisfied by the T-dual supervielbeins if and only if the right-hand side of~\eqref{eq:Lprojection} equals the expression $\tilde\bme^{\gamma(\alpha}\wedge\tilde\bmomega_\gamma{}^{\beta)}-\tfrac12\tilde\bme^\alpha\wedge\tilde\bme^\beta$. In particular, this implies that the terms in~\eqref{eq:Lprojection} proportional to $\tilde\bme^\alpha\wedge\tilde\bme^\beta$ have to sum up to a constant. From the expressions above, we see that the desired $\tilde\bme^\alpha\wedge\tilde\bme^\beta$ term is contained in $\upsilon^{\alpha\beta}$ defined in~\eqref{eq:ee}, which in turn appears in $Z^{\alpha\beta}$. Therefore, focusing only on the $\upsilon^{\alpha\beta}$ terms, we obtain
        \begin{equation}\label{eq:someResult}
            \begin{aligned}
                \rmd\tilde\bme^{\alpha\beta}\ &=\ -\frac12\upsilon^{\alpha\beta}-\frac{1}{2(1-2\tilde x^2)}\big[(1-\tilde x^2)\upsilon^{\alpha\beta}-2\tilde x_\gamma{}^\alpha\tilde x_\delta{}^\beta\upsilon^{\gamma\delta}-2\rmi\tilde x_\gamma{}^{(\alpha}\upsilon^{\beta)\gamma}\big]
                \\
                &\kern2cm-\frac{\rmi}{2(2-\tilde x^2)(1-2\tilde x^2)^2}\Big[(1+2\tilde x^4)\upsilon^{\alpha\beta}-4(2-\tilde x^2)\tilde x_\gamma{}^\alpha\tilde x_\delta{}^\beta\upsilon^{\gamma\delta}
                \\
                &\kern4cm-\rmi(5+2\tilde x^2)\tilde x_\gamma{}^{(\alpha}\upsilon^{\beta)\gamma}\Big]\tilde\theta^2+\cdots~,
            \end{aligned}
        \end{equation}
        where the ellipses denotes all the other terms that are irrelevant to the argument. Now, it is easy to realise that in~\eqref{eq:someResult} there are non-constant terms proportional to $\tilde\bme^\alpha\wedge\tilde\bme^\beta$ such as $\tilde\bme^\gamma\wedge\tilde\bme^\delta\tilde x_{(\gamma}{}^{(\alpha}\tilde x_{\delta)}{}^{\beta)}$. However, such terms can never sum up to a constant. Consequently, we may conclude that the T-dual supervielbeins $\tilde\bme$ do not satisfy the torsion constraints of three-dimensional supergravity. In order to improve the situation, we could consider performing a local $\sfOSp(1|2)$-transformation of the form $\tilde\bme\mapsto\tilde\bme^g\coloneqq g^{-1}\tilde\bme g$ for $g\in\scC^\infty(\Sigma,\sfOSp(1|2))$. Then, with $\Lambda^g\coloneqq g^{-1}\Lambda g$ and $A\coloneqq g^{-1}\rmd g$, the structure equation~\eqref{eq:MCTdualVielBeinPCM} becomes
         \begin{equation}\label{eq:gaugeTransformedTDualMCPCM}
            \nabla_A\tilde\bme^g\ =\ -\frac12[\tilde\bme^g,\tilde\bme^g]-\frac12\frac{1}{1-\ad_{\Lambda^g}}([\tilde\bme^g,\tilde\bme^g])~.
        \end{equation}
        However, one quickly realises that whilst such transformations can in principle remove some of the unwanted terms in~\eqref{eq:someResult}, they will never remove terms of the form $\tilde\bme^\gamma\wedge\tilde\bme^\delta\tilde x_{(\gamma}{}^{(\alpha}\tilde x_{\delta)}{}^{\beta)}$ simply because $A$ appears linearly in~\eqref{eq:gaugeTransformedTDualMCPCM}.

        An additional confirmation of the incompatibility of the dual model with supergravity comes from considering the three-form curvature $\tilde H_3$, which should have the particular structure~\eqref{eq:H3kuzenko} in terms of an arbitrary scalar superfield. Using the formula~\eqref{eq:dualPCMH3} for $\tilde H_3$ together with~\eqref{eq:someAdAction}, it is not too difficult to find the explicit expression of $\tilde H_3$ generalising~\eqref{eq:dualSL2H3} to the supersetting. In particular, it turns out that it contains a non-constant term which is cubic in the fermionic components of the supervielbein,
        \begin{equation}\label{eq:H-full-dualisation}
            \tilde H_3\ =\ \tilde\bme^\alpha\wedge\tilde\bme^\beta\wedge\tilde\bme^\gamma\frac{2}{1-2\tilde x^2}\bigg(\rmi\tilde x_{(\alpha\beta}\tilde\theta_{\gamma)}-\frac{3}{2-\tilde x^2}\tilde x_{(\alpha\beta}\tilde x_{\gamma)\delta}\tilde\theta^\delta\bigg)+\cdots~.
        \end{equation}
        Such a term is not present in~\eqref{eq:H3kuzenko}. This again shows that the T-dual model is incompatible with the supergravity constraints.

        One may wonder whether this result is a peculiarity of dualising all of $\sfOSp(1|2)$. In \cref{app:SL2-dualisation}, we comment on the maximal bosonic subgroup $\sfSL(2,\IR)\subseteq\sfOSp(1|2)$ and argue that also in this case the dualisation is not compatible with the supergravity constraints.

        The above issue may be related to the fact that our starting point was not a three-dimensional Green--Schwarz superstring sigma model on AdS$_3$, while the fermionic non-Abelian T-duality of stringy sigma models was argued in~\cite{Borsato:2017qsx,Borsato:2018idb} to produce dual superbackgrounds that satisfy supergravity constraints. The AdS$_3$ superstring sigma model is obtained by subtracting from the action~\eqref{eq:osp-initial-action} the term $j^\alpha\wedge{\star j_\alpha}$ and adding, instead of it, the Wess--Zumino term associated with the worldvolume pull-back of $H_3$ in~\eqref{eq:H3ads3} with a relative coefficient $k$ that ensures kappa-symmetry,
        \begin{equation}\label{eq:GSstring}
            S_\text{GS}\ =\ -\tfrac12\int_\Sigma\,\bme^{\alpha\beta}\wedge{\star\bme_{\alpha\beta}}+k\,\int_{M_3}\,H_3
            \ewith
            \partial M_3\ =\ \Sigma~.
        \end{equation}
        In the presence of such Wess--Zumino terms the gauging procedure is however more delicate~\cite{Hull:1989jk} and the approach we took in this work would not be applicable. Note that the three-dimensional superstring sigma model under consideration is similar to ten-dimensional superstrings in an Neveu--Schwarz--Neveu--Schwarz background which are not described by sigma models on semi-symmetric Ramond--Ramond superbackgrounds. The $B_2$-field of the AdS$_3$ superstring is invariant under the super non-Abelian isometry only modulo a gauge transformation, and, as such, it falls out of the class of examples considered in~\cite{Borsato:2018idb}. For this reason, the dualisation of the three-dimensional Green--Schwarz superstring deserves further study. Another possibility of tackling this issue might be to generalise the above three-dimensional superspace construction to a type IIB Green--Schwarz superstring in an AdS$_3\times S^3\times{\rm CY}_2$ Ramond--Ramond background (where ${\rm CY}_2$ is a Calabi--Yau two-fold). This would require the consideration of a semi-symmetric space sigma model with an isometry supergroup $\sfPSU(1,1|2)\times\sfPSU(1,1|2)$ rather than a higher-dimensional supergroup of the $\sfOSp$ series. The non-Abelian T-dualisation of the $\sfSL(2,\IR)$ isometry subgroup in the bosonic string sigma model on this background was recently considered in~\cite{Ramirez:2021tkd}, where it was shown that a dual background AdS$_2\times\IR\times S^3\times{\rm CY}_2$ is a solution of massive type IIA supergravity. It would certainly be of interest to generalise this example to its super-non-Abelian T-dualised counterpart, and we postpone the study of this problem to future work.

        \section{Super non-Abelian T-duality of supercoset models}\label{sec:cosets}

        We shall now move on and discuss the super non-Abelian T-duality procedure for symmetric and semi-symmetric spaces from \cref{sec:Symmetric,sec:Semisymmetric}. As indicated, the dualisation procedure for sigma models on coset superspaces $\sfG/\sfH$ is affected by additional technical complications related to certain uniqueness issues when trying to solve the equation of motion for $\omega$ for the projection of $j_\omega$ onto $\frh$. This is simply due to the fact that the action for coset models does not involve the projection of $j$ onto $\frh$. In addition, the non-invertibility can also be brought into the game by the fermionic coordinates which are intrinsically non-invertible. This peculiarity prevents us from presenting the super non-Abelian T-duality procedure in complete generality, as a case-by-case analysis would be necessary. For this reason, we shall only present the general approach, in particular pointing out where the non-invertibility arises, and postpone to future work the study of specific models of physical relevance. Nevertheless, we can still discuss various properties of the T-dual model such as integrability and the exchange of the equations of motion with the Maurer--Cartan equations which is an identifier pattern of dualisation.

        \subsection{Symmetric space sigma models}\label{sec:TDualSymmetricSpace}

        \paragraph{Gauged sigma model action.}
        Starting point is the action~\eqref{eq:symmetricSpaceAction}. Upon gauging a subgroup $\sfK\subseteq\sfG$, we decompose the covariant current~\eqref{eq:gaugedFlatCurrent} as
        \begin{equation}\label{eq:currentdecomposition}
            \begin{gathered}
                j_\omega\ =\ A_\omega+m_\omega~,
                \\
                A_\omega\ \coloneqq\ P_\frh(j_\omega)\ =\ A+P_\frh(g^{-1}\omega g)
                \eand
                m_\omega\ \coloneqq\ P_\frm(j_\omega)\ =\ m+P_\frm(g^{-1}\omega g)~,
            \end{gathered}
        \end{equation}
        where $A$ and $m$ were introduced in~\eqref{eq:symmetricSpaceFlatCurrent} and the projectors $P_\frh$ and $P_\frm$ in~\eqref{eq:symmetricSpaceDecomposition}, respectively. Consequently, the master action~\eqref{eq:MasterAction2} becomes
        \begin{equation}\label{eq:gaugedSymmetricSpaceAction}
            S_\omega\ =\ \tfrac12\int_\Sigma\inner{m_\omega}{\star m_\omega}+\tfrac12\int_\Sigma\inner{D(j_\omega)}{j_\omega}+\int_\Sigma\inner{\tilde\Lambda}{F_{j_\omega}}~.
        \end{equation}

        \paragraph{T-dual sigma model action.}
        The variation of~\eqref{eq:gaugedSymmetricSpaceAction} with respect to $\omega$ yields
        \begin{subequations}\label{eq:EOMSymmetricSpaceGauged}
            \begin{equation}
                {\star m_\omega}+\nabla_{j_\omega}\tilde\Lambda-D(j_\omega)\ =\ 0~.
            \end{equation}
            Upon decomposing $\tilde\Lambda$ as $\tilde\Lambda=\tilde\Lambda_\frh+\tilde\Lambda_\frm$ with $\tilde\Lambda_\frh\in\scC^\infty(\Sigma,\frh)$ and $\tilde\Lambda_\frm\in\scC^\infty(\Sigma,\frm)$, renaming $\tilde A\coloneqq A_\omega$, using~\eqref{eq:definitionCovariantD}, and recalling that $D:\frh\leftrightarrow\frm$, we can express~\eqref{eq:EOMSymmetricSpaceGauged} as
            \begin{equation}\label{eq:EOMSymmetricSpaceGaugedb}
                D_{\tilde\Lambda_\frm}(m_\omega)\ =\ \nabla_{\tilde A}\tilde\Lambda_\frh
                \eand
                m_\omega-[\tilde\Lambda_\frh,{\star m_\omega}]\ =\ -{\star\big(\rmd\tilde\Lambda_\frm-D_{\tilde\Lambda_\frm}(\tilde A)\big)}\,.
            \end{equation}
        \end{subequations}
        Next, with the help of the projectors~\eqref{eq:ASDProjectors}, the second equation in~\eqref{eq:EOMSymmetricSpaceGaugedb} is solved for $m_\omega$ as
        \begin{equation}\label{eq:solTdualSymmetricSpaceCurrent}
            m_\omega\ =\ -\frac{1}{1-\ad_{\tilde\Lambda_\frh}}\Big(P^+\big(\rmd\tilde\Lambda_\frm-D_{\tilde\Lambda_\frm}(\tilde A)\Big)+\frac{1}{1+\ad_{\tilde\Lambda_\frh}}\Big(P^-\big(\rmd\tilde\Lambda_\frm-D_{\tilde\Lambda_\frm}(\tilde A)\big)\Big)\,.
        \end{equation}
        Finally, the substitution of this expression into the action~\eqref{eq:gaugedSymmetricSpaceAction} yields
        \begin{equation}\label{eq:SymmetricDualAction}
            \tilde S\ =\ \int_\Sigma\innerLarge{\big(\rmd\tilde\Lambda_\frm-D_{\tilde\Lambda_\frm}(\tilde A)\big)}{\frac{1}{1-\ad_{\tilde\Lambda_\frh}}\Big(P^+\big(\rmd\tilde\Lambda_\frm-D_{\tilde\Lambda_\frm}(\tilde A)\Big)}+\int_\Sigma\inner{\tilde\Lambda_\frh}{F_{\tilde A}}~.
        \end{equation}
        Evidently, we have not yet used the first equation of~\eqref{eq:EOMSymmetricSpaceGaugedb}, and, as such, it is a hybrid action. To obtain the fully T-dualised action, we would have to substitute~\eqref{eq:solTdualSymmetricSpaceCurrent} into this equation to obtain a linear equation for $\tilde A$. However, generically, the resulting equation will not admit a unique solution for $\tilde A$ as the linear operator one would have to invert may have a non-trivial kernel. This is the non-invertibility issue to which we have alluded before. Of course, in specific situations, there will be a unique solution, and in those cases we can substitute this solution into~\eqref{eq:SymmetricDualAction} to obtain the fully T-dualised action. Then, of the original coordinates, we can gauge-fix $n^\sfH_\ttb+n^\sfK_\ttb$ bosonic and $n^\sfH_\ttf+n^\sfK_\ttf$ fermionic ones so that the T-dual model describes the dynamics of $n^\sfG_\ttb-n^\sfH_\ttb$ bosonic and $n^\sfG_\ttf-n^\sfH_\ttf$ fermiomic degrees of freedom with $n^\sfG_\ttb-n^\sfH_\ttb-n^\sfK_\ttb$ and $n^\sfG_\ttf-n^\sfH_\ttf-n^\sfK_\ttf$ coming from original coordinates and $n^\sfK_\ttb$ and $n^\sfK_\ttf$ from the Lagrange multipliers, respectively.

        In the following, we shall ignore this non-invertibility issue and proceed with~\eqref{eq:SymmetricDualAction}. Note that the first equation of~\eqref{eq:EOMSymmetricSpaceGaugedb} arises from varying~\eqref{eq:SymmetricDualAction} with respect to $\tilde A$. Note also that an advantage of this hybrid action is that it enjoys an $\sfH$-gauge invariance by means of
        \begin{equation}
            \tilde A\ \mapsto\ h^{-1}\tilde A h+h^{-1}\rmd h
            \eand
            \tilde\Lambda\ \mapsto\ h^{-1}\tilde\Lambda h+h^{-1}D(h)
        \end{equation}
        for all $h\in\scC^\infty(\Sigma,\sfH)$, and this allows us to demonstrate the exchange of Maurer--Cartan equations and equations of motion momentarily. The transformation rule for $\tilde\Lambda$ follows directly from~\eqref{eq:MasterAction2} using~\eqref{eq:HAction}. Note that $h^{-1}D(h)\in\frm$ for all $h\in\scC^\infty(\Sigma,\sfH)$.

        \paragraph{T-dual equations of motion and Lax connection.}
        The difficulty in solving the equations of motion for the gauge connection does not prevent from studying some properties of the T-dual model in complete generality. One of these properties is classical integrability, that is the existence of a dual Lax connection.

        In particular, the variations of the action~\eqref{eq:SymmetricDualAction} with respect to $\tilde\Lambda_\frh$ and $\tilde\Lambda_\frm$ yield the  equations of motion
        \begin{subequations}\label{eq:EOMdual}
            \begin{equation}
                F_{\tilde A}+\tfrac12[\tilde m,\tilde m]\ =\ 0
                \eand
                \nabla_{\tilde A}\tilde m\ =\ 0~,
            \end{equation}
            where we have used~\eqref{eq:variation2} and~\eqref{eq:variation1} and defined $\tilde m\coloneqq m_\omega$ with $m_\omega$ as given in~\eqref{eq:solTdualSymmetricSpaceCurrent}. Furthermore, the variation with respect to $\tilde A$ leads to
            \begin{equation}
                D_{\tilde\Lambda_\frm}(\tilde m)\ =\ \nabla_{\tilde A}\tilde\Lambda_\frh~.
            \end{equation}
        \end{subequations}
        Taking into account that $\tilde m$ satisfies the second equation in~\eqref{eq:EOMSymmetricSpaceGaugedb} identically and using the equations of motion above and the Jacobi identity, after a short calculation we obtain
        \begin{equation}\label{eq:dualconstraint}
            \nabla_{\tilde A}{\star\tilde m}\ =\ 0~.
        \end{equation}
        A comparison between the structure of the dual equations of motion~\eqref{eq:EOMdual} and the constraint~\eqref{eq:dualconstraint} with the ones of the original model~\eqref{eq:EOMSymmetricSpace} and~\eqref{eq:MCSymmetricSpace}, leads to the conclusion that also for symmetric space sigma models super non-Abelian T-duality exchanges equations of motion with the Maurer--Cartan equations. This allows to use
        \begin{equation}
            \tilde J(\tilde z)\ \coloneqq\ \tilde A+\tfrac12(\tilde z^2+\tilde z^{-2})\,\tilde m-\tfrac12(\tilde z^2-\tilde z^{-2})\,{\star\tilde m}~.
        \end{equation}
        as the T-dual Lax connection which ensures integrability.

        \subsection{Semi-symmetric space sigma models}

        \paragraph{Gauged sigma model action.}
        Finally, we discuss super non-Abelian T-duality for semi-symmetric space sigma models on $\sfG/\sfH$. Since the procedure is exactly the same as the one for symmetric space sigma models we can be rather brief.

        In the conventions of \cref{sec:Semisymmetric}, upon gauging a Lie subsupergroup $\sfK\subseteq\sfG$, we decompose the current~\eqref{eq:gaugedFlatCurrent} as
        \begin{equation}\label{eq:currentdecomposition2}
            \begin{gathered}
                j_\omega\ =\ A_\omega+p_\omega+m_\omega+q_\omega~,
                \\
                A_\omega\ \coloneqq\ P_\frh(j_B)\ =\ A+P_\frh(g^{-1}\omega g)~,
                \quad
                m_\omega\ \coloneqq\ P_\frm(j_\omega)\ =\ m+P_\frm(g^{-1}\omega g)~,
                \\
                p_\omega\ \coloneqq\ P_\frp(j_\omega)\ =\ p+P_\frp(g^{-1}\omega g)~,
                \quad
                q_\omega\ \coloneqq\ P_\frq(j_\omega)\ =\ q+P_\frq(g^{-1}\omega g)~,
            \end{gathered}
        \end{equation}
        where $A$, $p$, $m$, and $q$ were introduced in~\eqref{eq:semiSymmetricSpaceFlatCurrent} and the projectors $P_\frh$, $P_\frp$, $P_\frm$, and $P_\frq$ in~\eqref{eq:semiSymmetricSpaceDecomposition}, respectively. Therefore, the master action~\eqref{eq:MasterAction2} becomes
        \begin{equation}\label{eq:gaugedSemiSymmetricSpaceAction}
            S_\omega\ =\ \tfrac12\int_\Sigma\inner{m_\omega}{\star m_\omega}+\tfrac12\int_\Sigma\inner{p_\omega}{q_\omega}+\tfrac12\int_\Sigma\inner{D(j_\omega)}{j_\omega}+\int_\Sigma\inner{\tilde\Lambda}{F_{j_\omega}}~.
        \end{equation}

        \paragraph{T-dual sigma model action.}
        Upon varying~\eqref{eq:gaugedSemiSymmetricSpaceAction} with respect to $\omega$, we obtain
        \begin{subequations}\label{eq:EOMSemiSymmetricSpaceGauged}
            \begin{equation}\label{eq:EOMSemiSymmetricSpaceGaugeda}
                {\star m_\omega}-\tfrac12p_\omega+\tfrac12q_\omega+\nabla_{j_\omega}\tilde\Lambda-D(j_\omega)\ =\ 0~.
            \end{equation}
            More explicitly, with $\tilde\Lambda=\tilde\Lambda_\frh+\tilde\Lambda_\frp+\tilde\Lambda_\frm+\tilde\Lambda_\frq$ and $\tilde\Lambda_\frh\in\scC^\infty(\Sigma,\frh)$, $\tilde\Lambda_\frp\in\scC^\infty(\Sigma,\frp)$, $\tilde\Lambda_\frm\in\scC^\infty(\Sigma,\frm)$, and $\tilde\Lambda_\frq\in\scC^\infty(\Sigma,\frq)$, renaming $A_\omega\coloneqq\tilde A$, using~\eqref{eq:definitionCovariantD} and recalling that $D:\frh\leftrightarrow\frm$ and $D:\frp\leftrightarrow\frq$, we have
            \begin{equation}\label{eq:EOMSemiSymmetricSpaceGauged2}
                \begin{aligned}
                    -[\tilde\Lambda_\frq,p_\omega]-D_{\tilde\Lambda_\frm}(m_\omega)-[\tilde\Lambda_\frp,q_\omega]\ &=\ -\nabla_{\tilde A}\tilde\Lambda_\frh~,
                    \\
                    -[\tilde\Lambda_\frh,p_\omega]-[\tilde\Lambda_\frq,m_\omega]-D_{\tilde\Lambda_\frm}(q_\omega)-\tfrac12p_\omega\ &=\ -\nabla_{\tilde A}\tilde\Lambda_\frp~,
                    \\
                    m_\omega-[\tilde\Lambda_\frp,{\star p_\omega}]-[\tilde\Lambda_\frh,{\star m_\omega}]-[\tilde\Lambda_\frq,{\star q_\omega}]\ &=\ -{\star\big(\rmd\tilde\Lambda_\frm-D_{\tilde\Lambda_\frm}(\tilde A)\big)}\,,
                    \\
                    -D_{\tilde\Lambda_\frm}(p_\omega)-[\tilde\Lambda_\frp,m_\omega]-[\tilde\Lambda_\frh,q_\omega]+\tfrac12q_\omega\ &=\ -\nabla_{\tilde A}\tilde\Lambda_\frq~.
                \end{aligned}
            \end{equation}
        \end{subequations}

        Next, we solve the second and fourth equations in~\eqref{eq:EOMSemiSymmetricSpaceGauged2} for $p_\omega$ and $q_\omega$ in terms of $m_\omega$. Specifically, if we define
        \begin{subequations}\label{eq:invertedRelationsSemiSymmetric}
            \begin{equation}\label{eq:Rmatrix}
                R\ \coloneqq\
                \begin{pmatrix}
                    -\frac{1}{1+c_+\circ D_{\tilde\Lambda_\frm}\circ\,c_-\circ D_{\tilde\Lambda_\frm}}\circ c_+ & -\frac{1}{1+c_+\circ D_{\tilde\Lambda_\frm}\circ\,c_-\circ D_{\tilde\Lambda_\frm}}\circ c_+\circ D_{\tilde\Lambda_\frm}\circ c_-
                    \\
                    -\frac{1}{1+c_-\circ D_{\tilde\Lambda_\frm}\circ\,c_+\circ D_{\tilde\Lambda_\frm}}\circ c_-\circ D_{\tilde\Lambda_\frm}\circ c_+ & \frac{1}{1+c_-\circ D_{\tilde\Lambda_\frm}\circ\,c_+\circ D_{\tilde\Lambda_\frm}}\circ c_-
                \end{pmatrix}
            \end{equation}
            with $c_\pm\coloneqq\frac{2}{1\pm2\ad_{\tilde\Lambda_\frh}}$, we can write
            \begin{equation}\label{eq:invertedRelationsSemiSymmetrica}
                \begin{pmatrix}
                    p_\omega
                    \\
                    q_\omega
                \end{pmatrix}
                \ =\ R
                \begin{pmatrix}
                    -\nabla_{\tilde A}\tilde\Lambda_\frp+\ad_{\tilde\Lambda_\frq}(m_\omega)
                    \\
                    -\nabla_{\tilde A}\tilde\Lambda_\frq+\ad_{\tilde\Lambda_\frp}(m_\omega)
                \end{pmatrix}.
            \end{equation}
            Upon subsituting these expressions into the third equation of~\eqref{eq:EOMSemiSymmetricSpaceGauged2} and using the projectors~\eqref{eq:ASDProjectors}, we can solve for $m_\omega$ to obtain
            \begin{equation}\label{eq:mSolZ4Case}
                m_\omega\ =\ \left(-\frac{1}{1-S}\circ P^++\frac{1}{1+S}\circ P^-\right)(T)
            \end{equation}
            with
            \begin{equation}\label{eq:SandT}
                \begin{aligned}
                    S\ &\coloneqq\ \ad_{\tilde\Lambda_\frh}+\big(\ad_{\tilde\Lambda_\frp},\ad_{\tilde\Lambda_\frq}\big)\circ R\circ
                    \begin{pmatrix}
                        0 & \id
                        \\
                        \id & 0
                    \end{pmatrix}
                    \circ
                    \begin{pmatrix}
                        \ad_{\tilde\Lambda_\frp}
                        \\
                        \ad_{\tilde\Lambda_\frq}
                    \end{pmatrix},
                    \\
                    T\ &\coloneqq\ \rmd\tilde\Lambda_\frm-D_{\tilde\Lambda_\frm}(\tilde A)+\Big(\big(\ad_{\tilde\Lambda_\frp},\ad_{\tilde\Lambda_\frq}\big)\circ R\Big)
                    \begin{pmatrix}
                        \nabla_{\tilde A}\tilde\Lambda_\frp
                        \\
                        \nabla_{\tilde A}\tilde\Lambda_\frq
                    \end{pmatrix}.
                \end{aligned}
            \end{equation}
        \end{subequations}
        We should now subsitute these expressions in the first equation in~\eqref{eq:EOMSemiSymmetricSpaceGauged2} to obtain a linear equation for $\tilde A$. However, as in the case of symmetric space sigma models, this equation will not, in general, admit a unique as the linear operator that needs to be inverted may have a non-trivial kernel. This would require a case-by-case analysis and so, we proceed with a hybrid formulation by only inserting the solutions for $m_\omega$, $p_\omega$, and $q_\omega$ into~\eqref{eq:gaugedSemiSymmetricSpaceAction}. We find
        \begin{equation}\label{eq:SemiSymmetricSpaceDualAction}
            \tilde S\ =\ \int_\Sigma\innerLarge{T}{\frac{1}{1-S}\big(P^+(T)\big)}+\frac12\int_\Sigma\innerLarge{
            \begin{pmatrix}
                \nabla_{\tilde A}\tilde\Lambda_\frq
                \\
                \nabla_{\tilde A}\tilde\Lambda_\frp
            \end{pmatrix}
            }{R
            \begin{pmatrix}
                \nabla_{\tilde A}\tilde\Lambda_\frp
                \\
                \nabla_{\tilde A}\tilde\Lambda_\frq
            \end{pmatrix}
            }+\int_\Sigma\inner{\tilde\Lambda_\frh}{F_{\tilde A}}~.
        \end{equation}
        The first equation in~\eqref{eq:EOMSemiSymmetricSpaceGauged2} arises from varying this action with respect to $\tilde A$. Note that this action enjoys an $\sfH$-gauge invariance by means of
        \begin{equation}
            \tilde A\ \mapsto\ h^{-1}\tilde A h+h^{-1}\rmd h
            \eand
            \tilde\Lambda\ \mapsto\ h^{-1}\tilde\Lambda h+h^{-1}D(h)
        \end{equation}
        for all $h\in\scC^\infty(\Sigma,\sfH)$. As before, the transformation rule for $\tilde\Lambda$ follows directly from~\eqref{eq:MasterAction2} using~\eqref{eq:HAction}. Note again that $h^{-1}D(h)\in\frm$ for all $h\in\scC^\infty(\Sigma,\sfH)$.

        \paragraph{T-dual equations of motion and Lax connection.}
        A lengthy but straightforward calculation reveals that the variations of the action~\eqref{eq:SemiSymmetricSpaceDualAction} with respect to $\tilde\Lambda_\frh$, $\tilde\Lambda_\frp$, $\tilde\Lambda_\frm$, and $\tilde\Lambda_\frq$ yield the equations of motion\footnote{For details on the derivation of these equations we refer the reader to \cref{app:dualaction}.}
        \begin{subequations}\label{eq:dualEOMsemisymmetric}
            \begin{equation}
                \begin{aligned}
                    F_{\tilde A}+\tfrac12[\tilde m,\tilde m]+[\tilde p,\tilde q]\ &=\ 0~,
                    \\
                    \nabla_{\tilde A}\tilde p+[\tilde m,\tilde q]\ &=\ 0~,
                    \\
                    \nabla_{\tilde A}\tilde m+\tfrac12[\tilde p,\tilde p]+\tfrac12[\tilde q,\tilde q]\ &=\ 0~,
                    \\
                    \nabla_{\tilde A}\tilde q+[\tilde m,\tilde p]\ &=\ 0~,
                \end{aligned}
            \end{equation}
            where $\tilde m\coloneqq m_\omega$, $\tilde p\coloneqq p_\omega$, and $\tilde q\coloneqq q_\omega$ with $m_\omega$, $p_\omega$, and $q_\omega$ as given  in~\eqref{eq:invertedRelationsSemiSymmetric}. Additionally, the variation with respect to $\tilde A$ yields
            \begin{equation}
                -[\tilde\Lambda_\frq,\tilde p]-D_{\tilde\Lambda_\frm}(\tilde m)-[\tilde\Lambda_\frp,\tilde q]\ =\ -\nabla_{\tilde A}\tilde\Lambda_\frh~.
            \end{equation}
        \end{subequations}
        Consequently, using this equation, the equations of motion~\eqref{eq:dualEOMsemisymmetric} and the fact that $\tilde m$, $\tilde p$, and $\tilde q$ satisfy the second, third, and fourth equation in~\eqref{eq:EOMSemiSymmetricSpaceGauged2} identically, and the Jacobi identity, some algebra shows that we also have
        \begin{equation}
            \begin{aligned}
                \nabla_{\tilde A}{\star\tilde m}-\tfrac12[\tilde p,\tilde p]+\tfrac12[\tilde q,\tilde q]\ &=\ 0~,
                \\
                [\tilde m,\tilde p+{\star\tilde p}]\ &=\ 0~,
                \\
                [\tilde m,\tilde q-{\star\tilde q}]\ &=\ 0~.
            \end{aligned}
        \end{equation}
        Hence, we again obtain the exchange of equations of motion and Maurer--Cartan equations under super non-Abelian T-duality. Therefore, the T-dual Lax connection is given by
        \begin{equation}
            \tilde J(\tilde z)\ \coloneqq\ \tilde A+\tilde z\tilde p+\tfrac12(\tilde z^2+\tilde z^{-2})\,\tilde m+\tilde z^{-1}\tilde q-\tfrac12(\tilde z^2-\tilde z^{-2})\,{\star\tilde m}~,
        \end{equation}
        which ensures the classical integrability of the T-dual model.

        \section{Conclusions}\label{sec:conclusions}

        We have considered a general procedure of non-Abelian T-duality of sigma models on group manifolds, symmetric, and semi-symmetric spaces with the emphasis on the T-dualisation of sigma models whose target spaces are supermanifolds possessing non-Abelian superisometries. All these models are classically integrable, and since super non-Abelian T-duality exchanges the equations of motion with Bianchi identities (i.e.~Maurer--Cartan equations), it is straightforward to write down the T-dual Lax connection. In general, super non-Abelian T-duality encounters a problem of integrating out all the variables of the initial model and getting the dual model described entirely with a dual set of coordinates.
        
        We have considered in full detail the relatively simple example of the $\sfOSp(1|2)$ principal chiral model in which this problem does not occur. We have, however, argued that even though the target superspace of the initial model is an AdS$_3$ solution of three-dimensional $\caN=1$  supergravity, the dual target superspace background does not satisfy the three-dimensional supergraity constraints. Such model cannot even be a solution of a generalised supergravity~\cite{Arutyunov:2015mqj,Wulff:2016tju} because also in that case the torsion supergravity constraints are the same. A reason for the breaking of such constraints may be related to the fact that the $\sfOSp(1|2)$ principal chiral model does not describe the dynamics of a string in three-dimensional $\caN=1$ supergravity. The $\caN=1$, $d=3$ superstring action is of the Green--Schwarz type (see e.g.~\cite{Bergshoeff:1985su}) with the Wess--Zumino term which in the case of the AdS$_3$ superspace is a three-dimensional counterpart of an Neveu--Schwarz--Neveu--Schwarz background of ten-dimensional superstrings. To the best of our knowledge, non-Abelian T-duality of string sigma models in Neveu--Schwarz--Neveu--Schwarz super-backgrounds has not been studied in the literature, and this may be an interesting problem to address in future work.
        
        Another point\footnote{We are grateful to the referee for bringing this point to our attention.} that deserves further investigation concerns the extension of the non-Abelian T-duality group defined in~\cite{Klimcik:1995ux,Hoare:2017ukq,Lust:2018jsx} to our super-non-Abelian-T-duality setting. Such a generalisation should exist as it would represent the non-Abelian analogue of the $\sfOSp(d_\ttb,d_\ttb|d_\ttf)$ action defined in~\cite{Fre:2009ki,Osten:2016dvf}. The latter extends the bosonic $\sfO(d_\ttb,d_\ttb)$ symmetry~\cite{Giveon:1991jj} to backgrounds with bosonic and fermionic Abelian isometries.

        \appendix
        \addappheadtotoc
        \appendixpage

        \section{Cocycles and derivations}\label{app:cocycles}

        Consider the derivation $D$ defined in~\eqref{eq:definitionD} and let $V\in\frg$. Then, using $\rme^{s\ad_V}(U)=\sum_{k=0}^\infty\frac{s^k}{k!}\ad_V^k(U)$, it is easy to see that
        \begin{equation}
            D(\rme^{\ad_V}(U))\ =\ \rme^{\ad_V}(D(U))+\sum_{k,l=0}^\infty\frac{1}{(k+l+1)!}(\ad^k_V\circ\ad_{D(V)}\circ\ad^l_V)(U)
        \end{equation}
        because of the derivation property of $D$. Next, since $\int_0^1\rmd t\,(1-t)^kt^l=\frac{k!l!}{(k+l+1)!}$, we can rewrite the sum as
        \begin{equation}
            \begin{aligned}
                \sum_{k,l=0}^\infty\frac{1}{(k+l+1)!}\ad^k_V\circ\ad_{D(V)}\circ\ad^l_V\ &=\ \int_0^1\rmd t\,\sum_{k,l=0}^\infty\frac{1}{k!l!}(1-t)^kt^l\ad^k_V\circ\ad_{D(V)}\circ\ad^l_V
                \\
                &=\ \int_0^1\rmd t\,\rme^{(1-t)\ad_V}\circ\ad_{D(V)}\circ\,\rme^{t\ad_V}~.
            \end{aligned}
        \end{equation}
        Consequently,
        \begin{equation}
            \begin{aligned}
                D(\rme^{\ad_V}(U))\ &=\ \rme^{\ad_V}\left(D(U)+\int_0^1\rmd t\,\Big(\rme^{-t\ad_V}\circ\ad_{D(V)}\circ\,\rme^{t\ad_V}\Big)(U)\right)
                \\
                &=\ \rme^{\ad_V}\left(D(U)+\left[\int_0^1\rmd t\,\rme^{-t\ad_V}(D(V)),U\right]\right)
                \\
                &=\ \rme^{\ad_V}\left(D(U)+\left[\sum_{k=0}^\infty\frac{(-1)^k}{(k+1)!}\ad^k_V(D(V)),U\right]\right)
                \\
                &=\ \rme^V\big(D(U)+\big[\rme^{-V}D(\rme^V),U\big]\big)\rme^{-V}~,
            \end{aligned}
        \end{equation}
        where we have inserted the definition~\eqref{eq:extensionD}.
        This verifies~\eqref{eq:AdActionD}.

        \section{Fermionic current}\label{app:fermionicCurrent}

        Let us derive~\eqref{eq:fermionicCurrent} which is a special case of a more general formula~\cite{Kallosh:1998zx,Claus:1998yw}. We consider the one-parameter family ($t\in\IR$)
        \begin{equation}
            j(t)\ \coloneqq\ \rme^{t\caF}j_\ttb\,\rme^{-t\caF}+\rme^{t\caF}\rmd\rme^{-t\caF}~.
        \end{equation}
        Consequently,
        \begin{equation}
            \partial_tj(t)\ =\ \rme^{t\ad_\caF}(-\nabla_{j_\ttb}F)~,
        \end{equation}
        where $\nabla_{j_\ttb}$ is as defined in~\eqref{eq:covariantDerivative}. Therefore,
        \begin{equation}
            j(t)\ =\ j_\ttb+\int_0^t\rmd s\,\rme^{s\ad_\caF}(-\nabla_{j_\ttb}\caF)
            \quad\Rightarrow\quad
            j_\ttf\ =\ \int_0^1\rmd s\,\rme^{s\ad_\caF}(-\nabla_{j_\ttb}\caF)
        \end{equation}
        since $j(t=1)=j$. Upon using $\rme^{s\ad_V}(U)=\sum_{k=0}^\infty\frac{s^k}{k!}\ad_V^k(U)$, we conclude that
        \begin{equation}\label{eq:fercur}
            \begin{aligned}
                j_\ttf\ &=\ -\sum_{k=0}^\infty\frac{1}{(2k+1)!}\ad_\caF^{2k}(\nabla_{j_\ttb}\caF)+\sum_{k=0}^\infty\frac{1}{(2k+2)!}\ad_\caF^{2k+1}(\nabla_{j_\ttb}\caF)
                \\
                &=\ -\frac{\sinh(\ad_\caF)}{\ad_\caF}(\nabla_{j_\ttb}\caF)-2\frac{\sinh^2(\frac12\ad_\caF)}{\ad_\caF}(\nabla_{j_\ttb}\caF)~.
            \end{aligned}
        \end{equation}
        This establishes~\eqref{eq:fermionicCurrent}.

        \section{Useful identities for \texorpdfstring{$\frosp(1|2)$}{osp(1|2)}}\label{app:Liealgebraidentities}

        Here we derive the set of identities that have been used in \cref{sec:OSp} for the $\sfOSp(1|2)$ principal chiral model. To this end, we introduce
        \begin{equation}
            U\ \coloneqq\ \underbrace{u^{\alpha\beta}L_{\alpha\beta}}_{\eqqcolon\,U_\ttb}+\underbrace{\chi^\alpha Q_\alpha}_{\eqqcolon\,U_\ttf}
            \eand
            V\ \coloneqq\ \underbrace{\,v^{\alpha\beta}L_{\alpha\beta}}_{\eqqcolon\,V_\ttb}+\underbrace{\eta^\alpha Q_\alpha}_{\eqqcolon\,V_\ttf}~.
        \end{equation}
        We wish to compute the action of the operator $\frac{1}{1-\ad_U}$ on $V$.

        We start by expanding the operator $\frac{1}{1-\ad_U}$ in powers of $\ad_{U_\ttf}$. Since  $\ad^k_{U_\ttf}=0$ for $k>2$, we obtain
        \begin{equation}\label{eq:expansion}
            \begin{aligned}
                \frac{1}{1-\ad_U}\ &=\ \frac{1}{1-\ad_{U_\ttb}}+\frac{1}{1-\ad_{U_\ttb}}\circ\ad_{U_\ttf}\circ\,\frac{1}{1-\ad_{U_\ttb}}
                \\
                &\kern1cm+\frac{1}{1-\ad_{U_\ttb}}\circ\ad_{U_\ttf}\circ\,\frac{1}{1-\ad_{U_\ttb}}\circ\ad_{U_\ttf}\circ\,\frac{1}{1-\ad_{U_\ttb}}~.
            \end{aligned}
        \end{equation}
        We then expand $\frac{1}{1-\ad_{U_\ttb}}$ in powers of $\ad_{U_\ttb}$ and apply it to $V_\ttb+V_\ttf$,
        \begin{equation}
            \frac{1}{1-\ad_{U_\ttb}}(V_\ttb+V_\ttf)\ =\ \sum_{k=0}^\infty\ad^k_{U_\ttb}(V_\ttb+V_\ttf)~.
        \end{equation}
        For the bosonic part we note that, defining $u^2$ as in~\eqref{eq:DefOfx2} and using~\eqref{eq:osp12CommutationRelations}, we obtain
        \begin{equation}
            \ad_{U_\ttb}^2(V_\ttb)\ =\ (u^2 v^{\alpha\beta}-2{u_\gamma}^\alpha{u_\delta}^\beta v^{\gamma\delta})L_{\alpha\beta}~.
        \end{equation}
        Consequently, acting recursively with $\ad_{U_\ttb}$, we find
        \begin{equation}\label{eq:XYbosonicCommutator}
            \begin{gathered}
                \ad_{U_\ttb}^{2k}(V_\ttb)\ =\ (2u^2)^{k-1}(u^2 v^{\alpha\beta}-2{u_\gamma}^\alpha{u_\delta}^\beta v^{\gamma\delta})L_{\alpha\beta}
                \efor
                k\ \geq\ 1~,
                \\
                \ad_{U_\ttb}^{2k}(V_\ttb)\ =\ \ad_{U_\ttb}(\ad_{U_\ttb}^{2k}(V_\ttb))=-2\rmi(2u^2)^k{u_\gamma}^\alpha v^{\beta\gamma}L_{\alpha\beta}
                \efor
                k\ \geq\ 0~.
            \end{gathered}
        \end{equation}
        For the fermionic part, a similar calculation leads to
        \begin{equation}
            \ad^{2k}_{U_\ttb}(V_\ttf)\ =\ \big(\tfrac12 u^2\big)^k\eta^\alpha Q_\alpha
            \eand
            \ad^{2k+1}_{U_\ttb}(V_\ttf)\ =\ -\rmi\big(\tfrac12 u^2\big)^k{u_\alpha}^\beta\eta^\alpha Q_\beta~.
        \end{equation}

        We are now ready to compute $\frac{1}{1-\ad_U}(V)$ using the expansion~\eqref{eq:expansion} and the previous identities. For the first term in the expansion we obtain
        \begin{equation}\label{eq:Xbos-on-Y}
            \begin{aligned}
                \frac{1}{1-\ad_{U_\ttb}}(V_\ttb+V_\ttf)\
                &=\ \frac{1}{1-2u^2}\big[(1-u^2) v^{\alpha\beta}-2{u_\gamma}^\alpha{u_\delta}^\beta v^{\gamma\delta}-2\rmi{u_\gamma}^{(\alpha}v^{\beta)\gamma}\big]L_{\alpha\beta}
                \\
                &\kern1cm+\frac{2}{2-u^2}\eta^\alpha({\delta_\alpha}^\beta-\rmi {u_\alpha}^\beta)Q_\beta
                \\
                &\eqqcolon\ v'^{\alpha\beta}L_{\alpha\beta}+\eta'^\alpha Q_\alpha~.
            \end{aligned}
        \end{equation}
        From this structure, it easily follows that
        \begin{equation}
            \left(\ad_{U_\ttf}\circ\frac{1}{1-\ad_{U_\ttb}}\right)(V)\ =\ \eta'^{(\alpha}\chi^{\beta)}L_{\alpha\beta}+\rmi v'^\alpha{}_\beta\chi^\beta Q_\alpha~.
        \end{equation}
        Simply repeating the same procedure a few more times, we obtain the second term in~\eqref{eq:expansion} according to
        \begin{equation}
            \begin{aligned}
                &\left(\frac{1}{1-\ad_{U_\ttb}}\circ\ad_{U_\ttf}\circ\frac{1}{1-\ad_{U_\ttb}}\right)(V)\ =
                \\
                &\kern2cm=\ \frac{1}{1-2u^2}\big[(1-u^2)\eta'^{(\alpha}\chi^{\beta)}-2{u_\gamma}^\alpha{u_\delta}^\beta\eta'^{(\gamma}\chi^{\delta)}-2\rmi{u_\gamma}^\alpha\eta'^{(\beta}\chi^{\gamma)}\big]L_{\alpha\beta}
                \\
                &\kern3cm+\frac{2}{2-u^2}\rmi v'^\alpha{}_\gamma\chi^\gamma({\delta_\alpha}^\beta-\rmi {u_\alpha}^\beta)Q_\beta
                \\
                &\kern2cm\eqqcolon\ v''^{\alpha\beta}L_{\alpha\beta}+\eta''^\alpha Q_\alpha~,
            \end{aligned}
        \end{equation}
        from which we easily find
        \begin{equation}
            \left(\ad_{U_\ttf}\circ\frac{1}{1-\ad_{U_\ttb}}\circ\ad_{U_\ttf}\circ\frac{1}{1-\ad_{U_\ttb}}\right)(V)\ =\ \eta''^{(\alpha}\chi^{\beta)}L_{\alpha\beta}+\rmi v''^\alpha{}_\beta\chi^\beta Q_\alpha~,
        \end{equation}
        and the third term
        \begin{equation}
            \begin{aligned}
                &\left(\frac{1}{1-\ad_{U_\ttb}}\circ\ad_{U_\ttf}\circ\frac{1}{1-\ad_{U_\ttb}}\circ\ad_{U_\ttf}\circ\frac{1}{1-\ad_{U_\ttb}}\right)(V)\ =
                \\
                &\kern2cm=\ \frac{1}{1-2u^2}\big[(1-u^2)\eta''^{(\alpha}\chi^{\beta)}-2{u_\gamma}^\alpha{u_\delta}^\beta\eta''^{(\gamma}\chi^{\delta)}-2\rmi{u_\gamma}^\alpha\eta''^{(\beta}\chi^{\gamma)}\big]L_{\alpha\beta}
                \\
                &\kern3cm+\frac{2}{2-u^2}\rmi v''^\alpha{}_\gamma\chi^\gamma({\delta_\alpha}^\beta-\rmi{u_\alpha}^\beta)Q_\beta~.
            \end{aligned}
        \end{equation}
        In summary,
        \begin{subequations}\label{eq:finalexpansion}
            \begin{equation}
                \begin{aligned}
                    \frac{1}{1-\ad_U}(V)\ &=\ \frac{1}{1-2u^2}\big[(1-u^2)Z^{\alpha\beta}-2{u_\gamma}^\alpha{u_\delta}^\beta Z^{\gamma\delta}-2\rmi{u_\gamma}^{(\alpha}Z^{\beta)\gamma}\big]L_{\alpha\beta}
                    \\
                    &\kern1cm+\frac{2}{2-u^2}\zeta^\alpha({\delta_\alpha}^\beta-\rmi {u_\alpha}^\beta)Q_\beta
                \end{aligned}
            \end{equation}
            with
            \begin{equation}
                \begin{aligned}
                    Z^{\alpha\beta}\ & \coloneqq\ v^{\alpha\beta}+(\eta'+\eta'')^{(\alpha}\chi^{\beta)}
                    \\
                    &\,=\ v^{\alpha\beta}+\frac{2}{2-u^2}\big[\eta^{(\alpha}-\rmi\eta^\gamma u_\gamma{}^{(\alpha}\big]\chi^{\beta)}
                    \\
                    &\kern1cm+\frac{\rmi}{(2-u^2)(1-2u^2)}\Big[\big(1-\tfrac12 u^2\big)v^{\alpha\beta}-3{u_\gamma}^\alpha{u_\delta}^\beta v^{\gamma\delta}-3\rmi{u_\gamma}^{(\alpha}v^{\beta)\gamma}\Big]\chi^2~,
                    \\
                    \zeta^\alpha\ & \coloneqq\ \eta^\alpha-\rmi(v'+v'')^{\alpha\beta}\chi_\beta
                    \\
                    &\,=\ \eta^\alpha-\frac{\rmi}{1-2u^2}\big[(1-u^2)u^{\alpha\beta}-2{u_\gamma}^\alpha{u_\delta}^\beta v^{\gamma\delta}-2\rmi{u_\gamma}^{(\alpha}v^{\beta)\gamma}\big]\chi_\beta
                    \\
                    &\kern1cm-\frac{\rmi}{(2-u^2)(1-2u^2)}\Big[\tfrac32\eta^\alpha-\rmi\big(\tfrac72-u^2\big)\eta^\beta u_\beta{}^\alpha\Big]\chi^2~.
                \end{aligned}
            \end{equation}
        \end{subequations}

        \section{Three-dimensional supergravity}\label{app:conformalSUGRA}

        \paragraph{Cartan structure equations.}
        Consider a $(p,q|2n)$-dimensional supermanifold $M$ with metric $g$ of Gra{\ss}mann degree zero. The structure group of $M$ is generically $\sfOSp(p,q|2n)$ which is the group preserving the canonical graded symmetric bilinear form on $\IR^{p,q|2n}$. In particular, we let $A,B,\ldots$ the $\IR^{p,q|2n}$-indices, and we denote the canonical graded-symmetric bilinear form on $\IR^{p,q|2n}$ by $\eta_{AB}$. We have $\eta_{AB}=(-1)^{|A||B|}\eta_{BA}$ where $|-|$ denotes the Gra{\ss}mann degree. Note that $\sfOSp(p,q|2n)$ is generated by $\frac12(p+q)(p+q-1)+n(2n+1)$ bosonic generators and $2(p+q)n$ fermionic generators.

        Let now $\bmE_A$ be the supervielbeins on $M$. The structure functions $f_{AB}{}^C$ are given by
        \begin{equation}
            [\bmE_A,\bmE_B]\ =\ f_{AB}{}^C\bmE_C~,
        \end{equation}
        where $[-,-]$ denotes the (graded) Lie bracket. Furthermore, we define the dual supervielbeins $\bme^A$ by $\bmE_A\intprod\bme^B=\delta_A{}^B$ where $\delta_A{}^B$ is the Kronecker symbol and `$\intprod$' denotes the interior product. We set $\delta^A{}_B=(-1)^{|A||B|}\delta_A{}^B$. Then, in the supervielbein basis we write
        \begin{equation}
            g\ =\ \tfrac12\bme^B\odot\bme^A\eta_{AB}\ =\ \bme^B\otimes\bme^A\eta_{AB}~,
        \end{equation}
        where `$\odot$' denotes the graded symmetric tensor product. The inverse $\eta^{AB}$ of $\eta_{AB}$ is given by
        \begin{equation}\label{eq:inverseMetric}
            \eta^{AC}\eta_{CB}\ =\ \delta^A{}_B
            \quad\Leftrightarrow\quad
            (-1)^{|C|}\eta_{AC}\eta^{CB}\ =\ \delta_A{}^B~.
        \end{equation}

        Let now `$\caL_X$' be the Lie derivative along a vector field $X$ on $M$ and `$\rmd$' the exterior derivative. Then, we have the standard Cartan formula $\caL_X\omega=\rmd(X\intprod\omega)+X\intprod\rmd\omega$ for any differential form $\omega$ on $M$, and since $\caL_X$ is a graded derivation with respect to the tensor product, these then imply that
        \begin{equation}
            \rmd\bme^A\ =\ \tfrac12\bme^C\wedge\bme^Bf_{BC}{}^A~.
        \end{equation}
        Here, `$\wedge$' denotes the graded antisymmetric tensor product.

        Next, we define the torsion and curvature two-forms,
        \begin{equation}
            \bmT^A\ =\ \tfrac12\bme^C\wedge\bme^B\bmT_{BC}{}^A
            \eand
            \bmR_A{}^B\ =\ \tfrac12\bme^C\wedge\bme^B\bmR_{BCA}{}^B~,
        \end{equation}
        by the Cartan structure equations
        \begin{equation}\label{eq:cartanStructureEquations}
            \rmd\bme^A-\bme^B\wedge\bmomega_B{}^A\ \eqqcolon\ -\bmT^A
            \eand
            \rmd\bmomega_A{}^B-\bmomega_A{}^C\wedge\bmomega_C{}^B\ \eqqcolon\ -\bmR_A{}^B~,
        \end{equation}
        where $\bmomega_A{}^B=\bme^C\bmomega_{CA}{}^B$ is the connection one-form.

        The Ricci tensor and the scalar curvature are then given by
        \begin{equation}
            \bmR_{AB}\ \coloneqq\ (-1)^{|B||C|}\bmR_{ACB}{}^C
            \eand
            \bmR\ \coloneqq\ \eta^{BA}\bmR_{AB}~,
        \end{equation}
        where $\eta^{AB}$ is the inverse metric as defined in~\eqref{eq:inverseMetric}.

        \paragraph{Connection one-form from metric compatibility.}
        The metric compatibility is expressed by
        \begin{equation}\label{eq:metricCompatibility}
            \bmomega_{AB}+(-1)^{|A||B|}\bmomega_{BA}\ =\ 0
            \ewith
            \bmomega_{AB}\ \coloneqq\ \bmomega_A{}^C\eta_{CB}~.
        \end{equation}
        Upon defining
        \begin{equation}
            \bmF_{AB}{}^C\ \coloneqq\ f_{AB}{}^C+\bmT_{AB}{}^C~,
        \end{equation}
        the Cartan structure equation~\eqref{eq:cartanStructureEquations} together with the metric compatibility~\eqref{eq:metricCompatibility} yield the components of the connection one-form as
        \begin{equation}
            \bmomega_{AB}{}^C\ =\ \tfrac12\big(\bmF^C{}_{AB}+(-1)^{|A||B|}\bmF^C{}_{BA}+\bmF_{AB}{}^C\big)\,,
        \end{equation}
        where $(-1)^{|C|(|A|+|B|)}\bmF^C{}_{AB}=\eta^{CD}\bmF_{DA}{}^E\eta_{EB}$.

        \paragraph{Three-dimensional supergravity constraints.}
        Let $M$ be a three-dimensional Lorentzian spin manifold. In that case, we have the factorisation $T_\IC M\cong S\odot S$ of the tangent bundle where $S$ is the spin bundle. This is equivalent to picking a conformal structure on $M$. In an orthonormal frame, the components of the metric on $M$ are $(\eta_{ab})=\diag(-1,1,1)$ with $a,b,\ldots=0,1,2$, and because of the identification $T_\IC M\cong S\odot S$, we may write $\eps_{\alpha(\gamma}\eps_{\delta)\beta}$ for $\eta_{ab}$ with the spinor indices $\alpha,\beta,\ldots=1,2$. Here, $\eps_{\alpha\beta}=-\eps_{\beta\alpha}$ is the standard symplectic structure on $S$ with $\eps_{\alpha\gamma}\eps^{\gamma\beta}=\delta_\alpha{}^\beta$ and $\delta_\alpha{}^\beta$ the Kronecker symbol.

        We are now interested in a $(1,2|2)$-dimensional supermanifold $M$ whose tangent bundle decomposes as $T_\IC M\cong S\odot S\oplus S[1]$ where $[1]$ denotes the Gra{\ss}mann-degree shift of the fibres. Hence, the generic structure group $\sfOSp(1,2|2)$ is reduced to $\sfSL(2,\IR)$ which is the three-dimensional tangent space Lorentz group. This is a general assumption for the form of supergravity geometry. Furthermore, we may decompose the index $A$ as $A=(\alpha\beta,\alpha)$ so that $\eta_{AB}$ is given by
        \begin{equation}
            (\eta_{AB})\ =\
            \begin{pmatrix}
                \eps_{\alpha(\gamma}\eps_{\delta)\beta} & 0
                \\
                0 & \rmi\eps_{\alpha\beta}
            \end{pmatrix}.
        \end{equation}
        Then, because of the identification $T_\IC M\cong S\odot S\oplus S[1]$, the components of the connection become related to each other
        \begin{subequations}\label{eq:3dConformalSUGRAconnection}
            \begin{equation}
                \bmomega_{\alpha\beta}{}^{\gamma\delta}\ =\ 2\bmomega_{(\alpha}{}^{(\gamma}\delta_{\beta)}{}^{\delta)}~,
            \end{equation}
            where $\bmomega_\alpha{}^\beta$ is the one-form spin connection on $S$. The metric compatibility implies that
            \begin{equation}\label{eq:3dMetricCompatibility}
                \bmomega_{\alpha\beta}\ =\ \bmomega_{\beta\alpha}
                \ewith
                \bmomega_{\alpha\beta}\ =\ \rmi\bmomega_\alpha{}^\gamma\eps_{\gamma\beta}~.
            \end{equation}
           In addition, to describe supergravity, one imposes on the supergeometry the following torsion constraints~\cite{Gates:1983nr,Buchbinder:1998qv,Buchbinder:2017qls}
            \begin{equation}\label{eq:3dConformalSUGRATorsion}
                \begin{aligned}
                    \rmd\bme^{\alpha\beta}-2\bme^{\gamma(\alpha}\wedge\bmomega_\gamma{}^{\beta)}\ &=\ -\tfrac12\bme^\alpha\wedge\bme^\beta~,
                    \\
                    \rmd\bme^\alpha-\bme^\beta\wedge\bmomega_\beta{}^\alpha\ &=\ -\tfrac12\bme^{\delta\epsilon}\wedge\bme^{\beta\gamma}\,\bmT_{\beta\gamma\,\delta\epsilon}{}^\alpha-\bme^\delta\wedge\bme^{\beta\gamma}\,\bmT_{\beta\gamma\,\delta}{}^\alpha~,
                \end{aligned}
            \end{equation}
            with no conditions on the torsion components $\bmT_{\alpha\beta\,\gamma\delta}{}^\epsilon$ and $\bmT_{\alpha\beta\,\gamma}{}^\delta$.
        \end{subequations}

        If three-dimensional $\caN=1$ supergravity contains a two-form superfield $B_2$, its three-form field strength $H_3$ is constrained in such a way that its components are functions of a scalar superfield $L$ and its derivatives (see e.g.~\cite{Buchbinder:2017qls}),
        \begin{equation}\label{eq:H3kuzenko}
            H_3\ =\ \bme_\alpha\wedge\bme^{\alpha\beta}\wedge\bme_\beta\,L+\bme_\beta\wedge\bme_\gamma{}^\beta\wedge\bme^{\gamma\alpha}\,\caD_\alpha L-\tfrac{\rmi}{6}\bme_\alpha{}^\beta\wedge\bme_\beta{}^\gamma\wedge\bme_\gamma{}^\alpha(\rmi\caD^2+8\caS)L
        \end{equation}
        where $\caD_\alpha$ is the superspace covariant derivative and $\caS$ is a scalar `prepotential' superfield which appears in the solution of the torsion constraints~\eqref{eq:3dConformalSUGRATorsion}.

        \section{T-dualisation of \texorpdfstring{$\sfSL(2,\IR)\subseteq\sfOSp(1|2)$}{SL(2,R) in OSp(1|2)}}\label{app:SL2-dualisation}

        Let us briefly discuss the dualisation of the maximal bosonic subgroup $\sfSL(2,\IR)\subseteq\sfOSp(1|2)$ of the $\sfOSp(1|2)$ principal chiral model. In particular, since $H^2(\frsl(2,\IR))=0$, we consider the T-dual action~\eqref{eq:PCMTdualAction} with $\tilde\Lambda=g^{-1}\Lambda g$ and $\Lambda\eqqcolon\tilde x^{\alpha\beta}L_{\alpha\beta}$. By construction, in this case,~\eqref{eq:PCMTdualAction} possesses an $\sfSL(2,\IR)$ gauge symmetry. Given the parametrisation~\eqref{eq:osp-parametrisation} for $g$, we can exploit this gauge invariance to set $g_\ttb=\unit$. Then, a short calculation reveals that
        \begin{equation}
            \tilde \Lambda\ =\ g_\ttf^{-1}\Lambda g_\ttf\ =\ \big(1+\tfrac\rmi4\theta^2\big)\tilde x^{\alpha\beta}L_{\alpha\beta}+\rmi\theta^\gamma\tilde x_\gamma{}^\alpha Q_\alpha\ \eqqcolon\ y^{\alpha\beta}L_{\alpha\beta}+\chi^\alpha Q_\alpha~.
        \end{equation}
        This differs from the $\tilde\Lambda$ used in the dualisation of $\sfOSp(1|2)$ in \cref{sec:OSp} by the changes $\tilde x^{\alpha\beta}\rightarrow y^{\alpha\beta}$ and $\tilde\theta^\alpha\rightarrow\chi^\alpha$. Therefore, we can rephrase the procedure given there by simply making these replacements. In particular, upon inspecting~\eqref{eq:someResult} once these replacements are made, it is evident that we cannot satisfy the first torsion constraint in~\eqref{eq:3dConformalSUGRATorsion}, exactly as in the case when dualising all of $\sfOSp(1|2)$. The same argument also applies to $H_3$ in~\eqref{eq:H-full-dualisation}.

        \section{Semi-symmetric space sigma model} \label{app:dualaction}

        Let us provide some more details on the T-dualisation of semi-symmetric space sigma models.

        \paragraph{T-dual action.}
        Firstly, let us explain as how to derive the T-dual action~\eqref{eq:SemiSymmetricSpaceDualAction}. In particular, using the decompositions $j_\omega=\tilde A+p_\omega+m_\omega+q_\omega$ and $\tilde\Lambda=\tilde\Lambda_\frh+\tilde\Lambda_\frp+\tilde\Lambda_\frm + \tilde\Lambda_\frq$ together with the fact that $D:\frh\leftrightarrow\frm$ and $D:\frp\leftrightarrow\frq$, we we can rewrite~\eqref{eq:gaugedSemiSymmetricSpaceAction} as
        \begin{equation}
            \begin{aligned}
                \tilde S\ &=\ \tfrac12\int_\Sigma\inner{m_\omega}{\star m_\omega}+\tfrac12\int_\Sigma\inner{p_\omega}{q_\omega}
                +\int_\Sigma\inner{\tilde\Lambda_\frh}{F_{\tilde A}+\tfrac12[m_\omega,m_\omega]+[p_\omega,q_\omega]}
                \\
                &\kern1cm+\int_\Sigma\inner{\tilde\Lambda_\frp}{\nabla_{\tilde A}q_\omega+[m_\omega,p_\omega]}+\int_\Sigma\inner{\tilde\Lambda_\frq}{\nabla_{\tilde A}p_\omega+[m_\omega,q_\omega]}
                \\
                &\kern2cm+\int_\Sigma\inner{\tilde\Lambda_\frm}{\nabla_{\tilde A}m_\omega+\tfrac12[p_\omega,p_\omega]+\tfrac12[q_\omega,q_\omega]}
                \\
                &\kern3cm+\int_\Sigma\inner{D(\tilde A)}{m_\omega}+\tfrac12\int_\Sigma\inner{D(p_\omega)}{p_\omega}+\tfrac12\int_\Sigma\inner{D(q_\omega)}{q_\omega}~.
            \end{aligned}
        \end{equation}
        Next, we replace $m_\omega$ by the equation of motion in~\eqref{eq:EOMSemiSymmetricSpaceGaugeda}, insert the second and fourth equations of motion in~\eqref{eq:EOMSemiSymmetricSpaceGauged2}, and successively use the explicit expressions of $p_\omega$, $q_\omega$, and $m_\omega$ given in~\eqref{eq:invertedRelationsSemiSymmetric}. Ultimately, this yields the expression~\eqref{eq:SemiSymmetricSpaceDualAction} for the dualised action.

        \paragraph{T-dual equations of motion.}
        In order to compute the equations of motion~\eqref{eq:dualEOMsemisymmetric}, we recall the definition of the matrix $R=(R_{ij})$ from~\eqref{eq:Rmatrix} and the operator $S$ from~\eqref{eq:SandT}, respectively. A quick calculation shows that
        \begin{subequations}\label{eq:identities-RandS}
            \begin{equation}\label{eq:Radjoint}
                \begin{gathered}
                    \inner{R_{12}(U)}{V}\ =\ -\inner{U}{R_{12}(V)}~,
                    \quad
                    \inner{R_{21}(U)}{V}\ =\ -\inner{U}{R_{21}(V)}~,
                    \\
                    \inner{R_{11}(U)}{V}\ =\ -\inner{U}{R_{22}(V)}~,
                \end{gathered}
            \end{equation}
            and therefore,
            \begin{equation}
                \innerLarge{\frac{1}{1\pm S}(U)}{V}\ =\ \innerLarge{U}{\frac{1}{1\mp S}(V)}
            \end{equation}
        \end{subequations}
        for all $U,V\in\Omega^p(\Sigma,\frg)$. Therefore, applying the general relation $\delta\big(\frac{1}{1\pm S}\big)=\mp\frac{1}{1\pm S}\circ\delta S\circ\frac{1}{1\pm S}$, the variation of the first term in~\eqref{eq:SemiSymmetricSpaceDualAction} is
        \begin{equation}
            \delta\innerLarge{T}{\frac{1}{1-S}\big(P^+(T)\big)}\ =\ -\inner{\delta T}{\tilde m}-\innerLarge{\bigg(\delta S\circ\frac{1}{1-S}\bigg)(T)}{\frac{1}{1+S}\big(P^-(T)\big)},
        \end{equation}
        where $\tilde m\coloneqq m_\omega$ with $m_\omega$ given in~\eqref{eq:mSolZ4Case}. Upon inspecting the definitions of $S$ and $T$ given in~\eqref{eq:SandT}, the only non-trivial part when computing $\delta S$ and $\delta T$ is the variation of $R$. Some algebra shows that the only non-vanishing variations of $R$ are
        \begin{equation}
            \delta_{\tilde\Lambda_\frh}R\ =\ R\circ\ad_{\delta\tilde\Lambda_\frh}\circ R
            \eand
            \delta_{\tilde\Lambda_\frm}R\ =\ R\circ
            \begin{pmatrix}
                0 & \id
                \\
                \id & 0
            \end{pmatrix}
            \circ\ad_{\delta\tilde\Lambda_\frm}\circ R~.
        \end{equation}
        Using all these formul{\ae}, it is now not too hard to derive the equations of motion~\eqref{eq:dualEOMsemisymmetric}.

    \end{body}

\end{document}